\documentclass[prc,twocolumn,showpacs,superscriptaddress]{revtex4-1}
\usepackage{dsfont}
\usepackage{graphicx,amsmath}
\def\be{\begin{eqnarray}}
\def\ee{\end{eqnarray}}
\def\bc{\begin{center}}
\def\ec{\end{center}}

\def\om{\omega}

\newcommand{\lsim}{\stackrel{\scriptstyle <}{\phantom{}_{\sim}}}

\begin{document}
\title{S-wave pion condensation in symmetric nuclear matter}
\author{D. N. Voskresensky}
 \affiliation{BLTP, Joint Institute for Nuclear Research, RU-141980 Dubna, Russia}\affiliation{ National Research Nuclear
    University (MEPhI),  Kashirskoe shosse 31,
    115409 Moscow, Russia}
\begin{abstract}
S-wave pion-nucleon interactions in the  linear sigma model, and in Manohar-Georgi  and Gasser-Sainio-Svarc models with finite number of terms in Lagrangians, as well as in a general phenomenological approach are reviewed. Subtleties associated with the current algebra theorems and field redefinitions are discussed. In the first and third models most likely the
s-wave pion condensation in the isospin-symmetric matter does not occur at least up to high densities, whereas within the second  model  it may appear already at moderate densities.  In the phenomenological approach two parameterizations of the s-wave pion-nucleon scattering amplitude and the pion  polarization operator  used in the literature are considered. The first parameterization employs  the off-mass-shell amplitude and allows to fulfil the current algebra theorems.
Using it the s-wave pion polarization operator in the isospin-symmetric matter is  reconstructed within the gas approximation. With  this pion polarization operator  the
s-wave pion condensation in the isospin-symmetric matter does not occur at least up to high densities. Second parameterization uses the on-mass-shell pion-nucleon scattering amplitude and does not satisfy the Adler and Weinberg conditions. With such a parameterization most likely  the s-wave pion condensation  in the isospin-symmetric matter may occur already at the nucleon density  $n\simeq (1.4-2.5) n_0$,
where $n_0$ is the density of the atomic nucleus, that should result in  observable effects.
Both parameterizations  allow to successfully describe the pion atom data.
\end{abstract}
\date{\today}
 \maketitle

\section{Introduction}\label{Intro}
Intensive study of pion degrees  of freedom in nuclear matter started in 1970s \cite{Migdal:1971cu,Migdal:1972,Scalapino:1972fu,Sawyer:1972cq,Migdal:1973bwk,
  Migdal:1973PL,Migdal:1973PL,Migdal:1973jkf,Migdal:1973zm,Baym:1973zk,Migdal:1974jn,
  Campbell:1974qt,Campbell:1974qu,Baym:1975tm}, cf. reviews \cite{Brown:1975di,Migdal:1978az,Voskresensky:1982vd,Ericson:1988gk,Migdal:1990vm} and further references therein. Frequency-momentum ($\omega, \vec{q}$) dependent pion  polarization operator and pion spectra were constructed for nuclear matter with arbitrary  ratio of neutrons and protons, $N/Z$,  and a possibility of the p-wave pion condensation in the baryon medium at the  density $n>n_c>n_0\simeq 0.5 m_{\pi}^3$ was demonstrated, where $n_0$ is the nuclear saturation density, $m_\pi$ is the pion mass.

  Pion off-mass-shell effects are of primary importance for description of the pion spectra and  a possibility of the pion condensation  \cite{Migdal:1972,Migdal:1974jn,Migdal:1978az}. The most important contribution to the pion polarization operator is given by the p-wave pion-nucleon and pion-$\Delta$-isobar interactions. For the s-wave contribution to $\pi^-$ polarization operator, $\Pi_S^{-}$,  the first works
   \cite{Migdal:1974jn,Baym:1975tm} employed  expression $\Pi_S^{\rm WT,-} \simeq C (n_n-n_p) \omega/m_\pi^2$,  with constant $C$ estimated as  $C\simeq (1-1.4)$,  $n_n$ and $n_p$ being the neutron and proton densities, respectively. Such a contribution is usually called the Weinberg-Tomozawa term. After replacements $\omega \to -\omega$, $\vec{q}\to -\vec{q}$  the same polarization operator describes $\pi^+$.  For   the case of the isospin-symmetric matter the mentioned  works  used  $\Pi_S (N\simeq Z)=0$ taking into account that  the experimental value of the   pion-nucleon scattering length is tiny. Papers \cite{Campbell:1974qt,Campbell:1974qu,Baym:1975tm,Voskresensky:1982vd} studied the problem of the description of  pion degrees of freedom in nuclear matter within linear sigma model as the realization of  approximate chiral symmetry with partial conservation of axial-vector current (PCAC), as canonical operator equation. Within this consideration  $C\simeq m^2_\pi/(2f^2_\pi)$, $f_\pi \simeq 92.4$ MeV is the weak pion decay constant.

 First phenomenological optical potential for description of the scattering of on-mass-shell mesons off nuclei   was suggested by L. S. Kisslinger in 1955 \cite{Kisslinger55}. Reference \cite{EE1966} introduced a phenomenological optical potential for the description  of pion atoms, cf. \cite{Ericson:1988gk}. Retardation effects were disregarded, whereas they actually play an important role in this problem, cf.  \cite{Migdal:1978az}. Appropriate fit of $\omega, \vec{q}$ dependent optical pion-nucleus potential to the pion-atom data  with $N=Z$ and $N\neq Z$ known to that time  was performed in \cite{Troitsky:1981}.

Parameterization of the optical pion-nucleus potential employed in \cite{Migdal:1978az,Troitsky:1981,Migdal:1990vm,Voskresensky:1993ud,Chanfray1994,Coon1999,Weise2005,Jido2008}
uses the fully off-mass-shell pion-nucleon amplitude, which  fulfills the current algebra theorems and the canonical PCAC condition.
 Oppositely, Refs. \cite{Delorme92,Ericson94,Kolomeitsev:2002gc,Kolomeitsev:2002mm,Hatsuda2010} used the on-mass-shell pion-nucleon amplitude, taking in-coming and out-going pion 4-momenta such that  $q^2=q^{\prime\,2}=m^2_\pi$, that  does not allow to fulfill  the so called Adler and Weinberg current algebra conditions.

The reasoning of \cite{Delorme92,Ericson94} and  their followers to put pions on mass shell considering amplitude of the pion-nucleon scattering and  the pion polarization operator in matter goes back to the equivalence theorem that  any  local change of variables in quantum field theories, which leaves the free field part of the Lagrangian unchanged, does not alter the $S$-matrix \cite{Haag1958,Chisholm1961,Kamefuchi1961}. Thereby  the models dealing  with the off-mass-shell amplitude (after variable replacement) and the on-mass-shell one,
being employed for description of purely  on-mass-shell particle scattering in vacuum,   should yield equivalent results.
The same statement holds for consideration of the  scattering of particles on a number of infinitely heavy centers \cite{AgassiGal}.  The authors \cite{Delorme92} spread these statements first to  the total amplitude of the pion scattering  on a system of massive (but not infinitely massive)  centers and then they   assumed that the theorem serves ``to fully eliminate  off-mass-shell effects both in the leading order and in higher order terms,'' even when one considers propagation  of the classical pion field in the medium. Further, one  concluded,  cf. \cite{Furnstahl2001,Furnstahl2002}, that off-mass-shell Green functions and all off-mass-shell properties are  unobservable.

However in reality the in-medium conserved current $j^\mu$ and energy momentum tensor $\Theta^{\mu\nu}$ are expressed in terms of the non-equilibrium  fully off-shell Green functions $G_C$ and self-energies $\Sigma_C$ determined on the Schwinger-Keldysh contour. In the thermal equilibrium they are further expressed through the spectral functions $A=-2\mbox{Im} G^R$, in-medium particle widths $\Gamma =-2\mbox{Im}\Sigma^R$  and the off-shell Fermi/Bose occupations $n_\omega$, where $G^R$ and $\Sigma^R$  are the retarded Green function and self-energy, cf. \cite{IKV-2,IKV1}. Conserved charges, the energy and the momentum are  observable quantities. Spectral functions $A=-2\mbox{Im} G^R$ (and flow $B$ and entropy flow $A_S$ spectral functions) are associated with the density of states and various time-delays, and in the virial limit with the measurable phase shifts, cf.   \cite{KVJPG40} and refs therein. Observable 3-momentum distributions of the particles radiated from a piece of a non-equilibrium matter are expressed in terms of  the non-equilibrium self-energy $i\Pi^{-+}$, the current-current correlator. The 3-momentum distributions of the particles radiated from a piece of the equilibrium matter are  expressed in terms of $A$, $\Gamma$ and $n_\omega$, cf. \cite{KV1996,VS87,SV89,Migdal:1990vm}. Particles radiated to infinity are on mass shell but all internal integrations are performed with the fully off-shell Green functions.
Authors  \cite{Furnstahl2002} suggest that the  3-momentum occupations in the medium, $\widetilde{n}_{\vec{k}}^{\rm med}$,  as  they are defined there, are not observable, since these quantities  depend on artificially introduced  interpolated fields in their example.  Here we should stress that   not the in-medium occupations $\widetilde{n}_{\vec{k}}^{\rm med}$
but various frequency integrals of $i\Pi^{-+}$, being calculated in a piece of matter, determine the  observable 3-momentum particle distributions at infinity, $n_{\vec{k}}^{\rm vac}$, particle luminosity, etc., cf. \cite{KVJPG40,KV2011}.
Moreover, the Noether and the in-medium 4-currents coincide only provided some special conditions are fulfilled, e.g. as it occurs  in the Fermi liquid theory and in so called $\Phi$ derivable approximation schemes, cf. \cite{IKV-2,IKV1}.
At the end let us note that the Landau damping, zero sound and phonon propagation, Landau-Pomeranchuk-Migdal effect,  phase transitions in condensed matter and many many other effects can be described only dealing with the off-mass-shell particle propagation.

Below it will be explicitly shown that the two  mentioned  (off-shell and on-shell) approaches  result in essentially different physical consequences,  that could  be checked experimentally. For example, extrapolation to densities higher than  the nuclear saturation density $n_0$ done within the model used in \cite{Migdal:1978az,Troitsky:1981,Migdal:1990vm,Voskresensky:1993ud} does not allow for the s-wave pion condensation in the isospin-symmetric nuclear matter, whereas the model employed in \cite{Delorme92,Ericson94,Kolomeitsev:2002gc,Kolomeitsev:2002mm} allows for  occurrence of the s-wave pion condensation already at the density  $n>(1.4-2.5) n_0$ or even for a smaller density, as it will be shown   below. Although the latter possibility was mentioned in \cite{Brown91}, Refs.
\cite{Delorme92,AdamiBrown93,Ericson94} disregarded it arguing that  a strong decrease of the effective pion mass  is compensated by the $\omega$-dependent range term in the spectrum, whereas, as it will be explicitly shown below, the possibility of the s-wave pion condensation  directly follows from  their model.

 Some works, cf.  \cite{BrownLeeRho,Kirchbach96}, tried to reconcile two mentioned approaches by doing formal replacements of the fields in the Lagrangians, which should not affect physics. The problem is  however subtle and some  authors changed their position from work to work, whereas in our opinion  solution of the puzzle is as follows:
 from the fact that observables should not depend on the choice of the interpolating fields in the {\em full Lagrangian} it does not follow that cancellation of artificially introduced  contributions depending on the interpolating fields should occur term by term or in the sub-groups of the diagrams. The graphs  should be calculated following the ordinary Feynman  rules, rather than by a somewhat artificial putting of the  in-going particles on mass shell in each diagram.  To keep in mind this circumstance it proves to be especially important in practical schemes, where one deals with the   Lagrangians, which differ at least in the high-order contributions in the fields (we  further call them {\em reduced Lagrangians}). Thereby it is not surprising that  such reduced Lagrangians   predict  different observable effects. Only experimental check of the specific predictions of the models  can allow to choose between them.

The paper is organized as follows. Next section formulates partial conservation of PCAC  and current algebra theorems. Then in section \ref{LinSigmaMod} we study conditions for their fulfilment  within the linear sigma model and then in section \ref{MonaharGeorgi}  and in section \ref{Gasser} we consider   the Manohar-Georgi  and the Gasser-Sainio-Svarc  reduced Lagrangians, respectively. {\em All models will be treated  at the usage of certain approximations.} We construct pion polarization operators in each of models. Section \ref{Phenomenological} discusses two purely phenomenological approaches to construct the s-wave pion-nucleon amplitude and the s-wave pion polarization operator in the isospin-symmetric nuclear matter. In the first  approach one deals with the pion-nucleon amplitude determined off mass shell at arbitrary relations between the  variables. The amplitude is constrained by the experimental fact that the s-wave pion-nucleon scattering length, $a^+_{\pi N}$, is very small and by
the fulfillment of the current algebra  Cheng-Dashen and  Weinberg (or   Adler) conditions. Correspondingly, the Adler (or Weinberg) condition is then fulfilled identically. In the second approach one puts  $q^2=q^{\prime\,2}=m^2_\pi$  satisfying the Cheng-Dashen condition and that $a^+_{\pi N}\simeq 0$ but violating the Adler and Weinberg conditions. Then section \ref{Tobe} focuses on a question about presence or absence  of the s-wave pion condensation in the isospin-symmetric nuclear matter. It will be shown that the s-wave pion condensation in the isospin-symmetric nuclear matter does not occur in the first approach and it may occur already at $n=(1.4-2.5)n_0$, or even at a smaller density, in the second approach. Section \ref{Conclusion} contains concluding remarks.

 \section{PCAC and current algebra}\label{PCACalgebra}

Low-energy pion-nucleon scattering was widely discussed in the 1960s, when ideas
of current algebra and PCAC were developed, cf. \cite{ChengDashen71,AdlerDashen68,Treiman72}.
One introduces  the isospin-even
pion-nucleon forward scattering amplitude with the pseudovector pole term subtracted,
\begin{eqnarray}
\widetilde{D}^{+}(\nu,t,q^2,q^{\prime 2})&
={D}^{+}(\nu,t,q^2,q^{\prime 2})
\label{sabtrAmpl}\\
&-\frac{\nu_B^2 g^2\Gamma (q^2)\Gamma (q^{\prime 2})}{m_N (\nu_B^2 -\nu^2)}\,,\nonumber
\end{eqnarray}
 where  $q^2=\omega^2 -\vec{q}^{\,2}$, $q^{\prime\,2}=\omega^{\prime 2}-\vec{q}^{\,\prime 2}$,   $\nu =(s-u)/(4m_N)=(p+p^{\prime})(q+q^{\prime})/(4m_N)$,  $\nu_B =(t-q^2-q^{\prime 2})/(4m_N)=-qq^{\prime}(2m_N)$
 , $s=(p+q)^2$, $u=(p^{\prime}-q)^2$, $t=(q-q^{\prime})^2$ are  appropriate  kinematical variables;
 $g$ and $\Gamma$ are, respectively, the $\pi NN$ coupling constant and a vertex form-factor, $m_N\simeq 938$ MeV is the nucleon mass in vacuum.

 The amplitude $\widetilde{D}^{+}$ is  related to the pion-nucleon sigma term, which is
an important measure of chiral symmetry breaking \cite{Pagels1971},
\begin{eqnarray}
\Sigma(t)=\frac{1}{3}\sum_{i=1}^{3} \langle N(p^{\prime}|[Q_5^i,[Q_5^i, H_{\rm SB}]]|N(p)\rangle\,,
\end{eqnarray}
where $N$ is the nucleon state, $H_{\rm SB}$ is the Hamiltonian density of the symmetry breaking term and
$Q^i_5=\int A_0^i(x)d^3x$ is the $i=1,2,3$ component of the axial-vector charge. On the quark level the axial current  is given by
$A^i_\mu =\bar{q}\gamma_\mu \gamma_5 \tau^i q$, where
$q$ are quark fields, $\gamma_\mu$,  $\gamma_5$ are Dirac matrices, $\tau^i$ are isospin matrices  in SU(2) case.
Recent lattice data \cite{Jang2020} satisfy PCAC  within 5 $\%$ errorbar.

Using the definition \cite{Lee72} of the pion decay constant $f_\pi$ for the process $\pi^-\to \mu^- +\bar{\nu}_\mu$ in vacuum, the hadronic matrix element of the axial current is
   \begin{eqnarray}
  \langle 0|A^i_\mu |\pi^j(q)\rangle \equiv iq_\mu f_\pi \delta^{ij}\,,
 \end{eqnarray}
that yields
\begin{eqnarray}
\langle 0|\partial^\mu A^i_\mu |\pi^j(q)\rangle =q^2 f_\pi \delta^{ij}=  m^2_\pi f_\pi \delta^{ij}\label{PCAC2}
\end{eqnarray}
for the pion in vacuum with $q^2 =m^2_\pi$.

Employing   convenient choice of the pion field operator normalization, $\langle 0|\pi^i|\pi^j\rangle =\delta^{ij}$, one gets
\begin{eqnarray}
\partial^\mu A^i_\mu =f_\pi m^2_\pi\pi^i  \,,\label{PCAC}
\end{eqnarray}
where $A^i_\mu$ is the hadronic axial vector current.

At the assumption of a smoothly varying amplitude $\tilde{D}^{+}(\nu,t,q^2,q^{\prime 2})$, one arrives at the relations \cite{ChengDashen71,AdlerDashen68} (in variables $\nu,\nu_B, q^2,q^{\prime 2}$ corresponding to $\nu=\nu_B=0$):
\begin{eqnarray}
\widetilde{D}^{+}(\nu=0,t=0,q^2=0,q^{\prime 2}=0)=-\Sigma (t=0)/f^2_\pi\label{Weinberg}
\end{eqnarray}
at the Weinberg   kinematical point;
\begin{eqnarray}
\widetilde{D}^{+}(0,m^2_\pi,m^2_\pi,0)=\tilde{D}^{+}(0,m^2_\pi,0,m^2_\pi)=0\label{Adler}
\end{eqnarray}
at the Adler   kinematical point;
and
\begin{eqnarray}
&\widetilde{D}^{+}(0,2m^2_\pi,m^2_\pi,m^2_\pi)= +\Sigma (t=2m^2_\pi)/f^2_\pi \nonumber\\
&\simeq
(\Sigma (t=0)/f^2_\pi) (1+O(m^2_\pi/m_{\sigma}^2))\label{ChengDashen}
\end{eqnarray}
at the Cheng-Dashen  kinematical point, cf. \cite{Campbell:1978,ThorssonWirsba95} and Eq. (\ref{sigmat}) below.
Here $m_{\sigma}$ is  the mass of the sigma meson.

The amplitude in the Weinberg point is repulsive and equal in magnitude to the
attractive amplitude in the Cheng-Dashen point. The Cheng-Dashen point is distinguished
from the others in the fact that both pions have their momenta on mass-shell, i.e.  $q^2=q^{\prime\,2}=m^2_\pi$. The sigma-commutator in (\ref{ChengDashen}) can be evaluated either at $t=2m^2_{\pi}$ or at $t = 0$, since the difference is a small correction, cf. \cite{Pagels1971}.

One also may assume that  putting the final pion back
on mass shell (and holding fixed $t$ and $\nu$) should not change the Adler consistency
condition much, cf. \cite{Coon1999}. If so, one gets an additional   condition
\begin{eqnarray}
\widetilde{D}^+ (0, m^2_\pi, m^2_\pi,0)\simeq
\widetilde{D}^+ (0, m^2_\pi, m^2_\pi,m^2_\pi)=0\,.\label{A2}
\end{eqnarray}

There exist various estimates of the pion-nucleon $\Sigma$-term in the literature, which cover a broad  range of values. Most modern estimates yield  $\Sigma\simeq 55-75$ MeV, cf. \cite{Dmitrasinovic16}. Using
the Roy-Steiner equations to control the extrapolation of the vanishingly
small near threshold $\pi N$ isoscalar scattering amplitude to zero pion mass, the  Bern-Bonn-J\"ulich group yielded $\Sigma\simeq 53-63$ MeV, cf.  \cite{Elvira18}.
The values of the  $\Sigma$ term extracted recently from a fit to the pion atoms \cite{FriedmanGal20}  correspond to $\Sigma\simeq 50-64$ MeV.

Reference \cite{Campbell:1978}  performed calculations in the tree
approximation to the linear sigma model. The canonical PCAC relation (\ref{PCAC}) and the  consistency conditions (\ref{Weinberg}), (\ref{Adler}), (\ref{ChengDashen}) are fulfilled at specific choices of symmetry breaking terms in the Lagrangian. Some Lagrangians, such as Manohar-Georgi one \cite{ManoharGeorgi84} including the next-to-leading order terms in the chiral perturbation
theory,  do not satisfy the canonical PCAC relation and (\ref{Weinberg}) and  (\ref{Adler}) conditions, cf.  discussion in  \cite{ThorssonWirsba95}.

\section{Linear sigma model}\label{LinSigmaMod}

\subsection{Lagrangian and PCAC }
The
Lagrangian density of the linear sigma model (SM)
is given by
\begin{eqnarray}
L_{\rm SM}=L^{\rm sym}_{\rm SM}+L^{\rm sb}\,,\label{linsigm}
\end{eqnarray}
where the symmetric part of the Lagrangian density (sym) is as follows
\begin{eqnarray}
&L^{\rm sym}_{\rm SM}=\bar{N}[i\gamma^\mu\partial_\mu -g(\sigma + i\vec{\tau}\vec{\pi}\gamma_5)]N\nonumber\\
&+[\partial_\mu \sigma \partial^\mu\sigma +\partial_\mu \vec{\pi}\partial^\mu \vec{\pi}]/2 -\lambda (\sigma^2+\vec{\pi}^{\,2}-v^2)^2/4\,.\label{linsigm1}
\end{eqnarray}
Here $\pi_i$ $\sigma$, $N$ are the isospin-vector pion, scalar sigma and bi-spinor neutron and proton fields and  $v$ is a positive constant. The symmetry breaking  term in the Lagrangian (sb) is taken as
\begin{eqnarray}
L^{\rm sb}=L^{\rm sb}_1+L^{\rm sb}_2+L^{\rm sb}_3=\epsilon_1\sigma -\epsilon_2 \vec{\pi}\vec{\pi}-\epsilon_3 \bar{N}N\,,\label{linsigm0}
\end{eqnarray}
 constants $\epsilon_i$ are assumed to be small quantities. Simplifying consideration we did not add terms responsible for the $\Delta$-isobar-pion-nucleon interaction. They also can be included within the SM, cf. \cite{Campbell:1974qt,Campbell:1974qu,Baym:1975tm}.

With the help of the axial transformations
\begin{eqnarray}
&\sigma\to \sigma +\alpha_i \pi_i\,,\quad  \pi_i\to \pi_i -\alpha_i \sigma\,,\nonumber\\
&N\to N+i\tau_i\alpha_i \gamma_5 N/2
\end{eqnarray}
from (\ref{linsigm1}) one finds expression for the axial-vector current
\begin{eqnarray}
A^i_\mu =\bar{N}\gamma^5\gamma_\mu\tau^iN/2+\pi^i\partial_\mu \sigma -\sigma \partial_\mu \pi^i\,.
\end{eqnarray}
This expression yields \cite{Lee72}
\begin{eqnarray}
\partial^\mu A^i_\mu =(\epsilon_1  +2\epsilon_2 \sigma)\pi^i -i\epsilon_3\bar{N}\gamma_5 \tau^i N\,.\label{PCAC1}
\end{eqnarray}
As we see, only for $\epsilon_2=\epsilon_3=0$, $\epsilon_1 \simeq m^2_\pi f_\pi$, the PCAC condition  holds in its canonical form (\ref{PCAC}). However in the tree approximation  it holds also for
\begin{eqnarray}
\epsilon_1  +2\epsilon_2\langle \sigma\rangle = m^2_\pi f_\pi\,,\label{treeAx}
\end{eqnarray}
other terms contribute to loops.
 The Adler consistency
condition is fulfilled provided $\epsilon_3 + 2\epsilon_2 m_N/m^2_\sigma =0$, see
Eq. (\ref{DoffAdler}) below.

\subsection{Mean field, particle masses and $\Sigma$ term}

 Up to linear terms in $\epsilon_i$ minimization  of the effective potential yields for the mean field $\langle\sigma\rangle$,
\begin{eqnarray}
\langle\sigma\rangle = v+\epsilon_1/(2\lambda v^2)+...\,,\label{sigmm}\end{eqnarray}
for the nucleon mass,
\begin{eqnarray}
m_N=g\langle\sigma\rangle +\epsilon_3\simeq gv+g\epsilon_1/(2\lambda v^2)+\epsilon_3+...\,,
\end{eqnarray}
  for the mass of the $\sigma^{\prime}= \sigma -\langle\sigma\rangle$ field counted from the expectation value,
 \begin{eqnarray}
 m^2_\sigma =\lambda (3\langle\sigma^2\rangle -v^2)\simeq 2\lambda v^2+3\epsilon_1/v+... \,,\label{sigmamm}
  \end{eqnarray}
 and   for the pion mass
  \begin{eqnarray}
  m^2_\pi =\lambda (\langle\sigma\rangle^2-v^2)+2\epsilon_2\simeq \epsilon_1/v +2\epsilon_2+...\,\label{pionmass}
  \end{eqnarray}

 Employing (\ref{PCAC2}), (\ref{PCAC1})
we find
\begin{eqnarray}
&\langle 0|(\epsilon_1+2\epsilon_2 \langle 0|\sigma|0\rangle )\pi^i+ (2\epsilon_2\sigma^{\prime}\pi^i -i\epsilon_3\bar{N}\gamma_5 \tau^i N)|\pi^j(q)\rangle\nonumber\\
&=f_\pi m^2_\pi \delta^{ij}\,.\label{langl}
\end{eqnarray}
In the tree approximation only first term contributes and we recover Eq. (\ref{treeAx}).
From Eqs. (\ref{treeAx}), (\ref{pionmass}) it follows that $\langle\sigma\rangle =f_\pi$.

For the $\Sigma$ term in the tree approximation Ref. \cite{Campbell:1978}
found expression
\begin{eqnarray}
&\Sigma (t)=\frac{(m_N-\epsilon_3)(m^2_\pi +2\epsilon_2)}{m^2_\sigma -t}+\epsilon_3\nonumber
\\
&\simeq\frac{m_N m^2_\pi}{m^2_\sigma}+\frac{ 2\epsilon_2 m_N}{m^2_\sigma}+\epsilon_3
\,
\label{sigmat}
\end{eqnarray}
and $\Sigma (t=2m^2_\pi)$ is close to $\Sigma (t=0)$ for $2m^2_\pi/m^2_\sigma\ll 1$.

With $m_N=939$ MeV, $m_\pi =139$ MeV and the ordinary used  value  $m_\sigma \simeq 600$ MeV
for $\epsilon_2=\epsilon_3=0$ we obtain $\Sigma (t=0)=m_N m^2_\pi/m^2_\sigma\simeq 50.4$ MeV, $\Sigma (t=2m^2_\pi)\simeq 56.5$ MeV, cf. (\ref{sigmat}).  These quantities are close to experimentally motivated values
\cite{Elvira18,FriedmanGal20}. We could consider it as an argument in favor of choosing
\begin{eqnarray}
\frac{2\epsilon_2 m_N}{m^2_\sigma}+\epsilon_3\simeq 0\,.\label{cond}
\end{eqnarray}
\begin{figure}\centering
\includegraphics[width=5.8cm,clip]{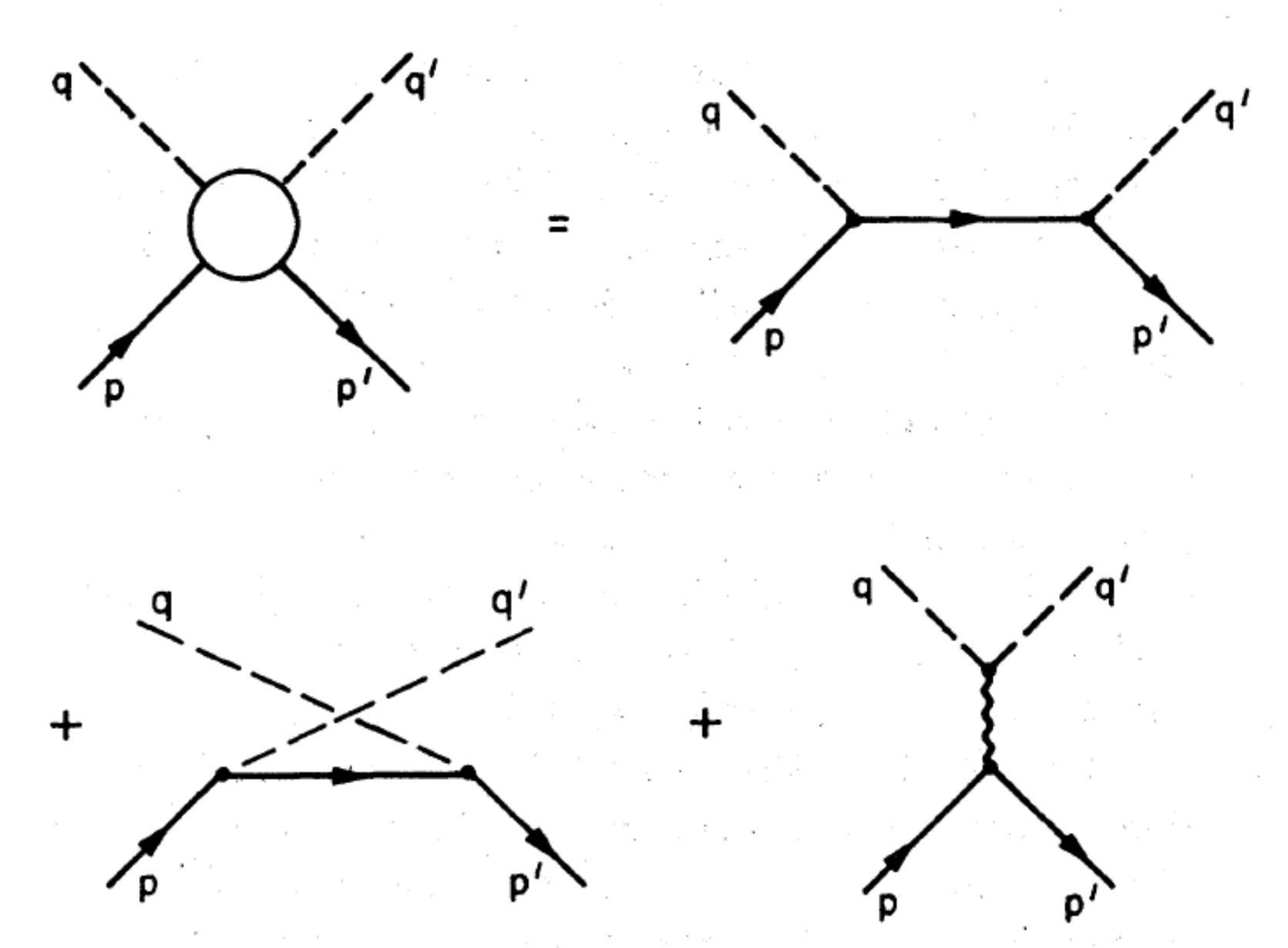}
\caption{Feynman diagrams contributing to $\pi N$ scattering amplitude in tree approximation to $\sigma$ model.
 }\label{PiNAmpl}
\end{figure}

\subsection{The $\pi N$ scattering amplitude}
Feynman diagrams contributing to the $\pi N$ scattering amplitude in the tree approximation to the $\sigma$ model \cite{Campbell:1978} are shown in Fig. \ref{PiNAmpl}.
First two diagrams  yield the standard pseudoscalar
Born pole
\begin{eqnarray}
{D}^+_{(1,2)}=\frac{g^2\nu^2}{m_N(\nu^2_{\rm B}-\nu^2)}
\end{eqnarray}
and the third diagram produces
\begin{eqnarray}
\widetilde{D}^+_{\rm (3)}=\frac{2\lambda v g}{m^2_\sigma -t}\,.
\end{eqnarray}
Replacing $\langle \sigma\rangle\simeq f_\pi$ and $2\lambda f_\pi\simeq (m^2_\sigma -m^2_\pi +2\epsilon_2)/f_\pi$, as it follows from (\ref{sigmm}), (\ref{sigmamm}), (\ref{pionmass}),  one finds
\begin{eqnarray}
\widetilde{D}^+=\frac{g^2}{m_N-\epsilon_3}-\frac{g^2}{m_N}+\frac{g^2}{m_N-\epsilon_3}
\frac{t-m^2_\pi +2\epsilon_2}{m^2_\sigma -t}\,,\label{Doff}
\end{eqnarray}
and in linear approximation in $\epsilon_i$,
\begin{eqnarray}
f^2_\pi \widetilde{D}^+\simeq \epsilon_3 +
\frac{m_N(t-m^2_\pi +2\epsilon_2)}{m^2_\sigma}\,.\label{Doffapp}
\end{eqnarray}

Putting $t=2m^2_\pi$ in this expression yields
\begin{eqnarray}
f^2_\pi\widetilde{D}^+ (t=2m^2_\pi)\simeq \frac{m_N m^2_\pi }{m^2_\sigma}
+\frac{2\epsilon_2 m_N}{m^2_\sigma}
+\epsilon_3 +...\label{ChD}
\end{eqnarray}
Comparing (\ref{ChD}) and (\ref{sigmat}) we obtain $f^2_\pi\widetilde{D}^+ (t=2m^2_\pi)\simeq \Sigma (0)\simeq \Sigma (t=2m^2_\pi)$. Thus Cheng-Dashen condition
(\ref{ChengDashen})
is fulfilled
provided $\Sigma (t=2m^2_\pi)\simeq \Sigma (t=0)=\Sigma$, i.e., when one neglects $2m^2_\pi \Sigma/m^2_\sigma\ll 1$ correction.

To obtain the amplitude in the Adler point we put
 $t=m^2_\pi$ and $q^{\prime}=0$  in (\ref{Doff}) that produces
\begin{eqnarray}
&\widetilde{D}^+ (0,m^2_\pi, m^2_\pi,0)=\frac{g^2}{m_N-\epsilon_3}-\frac{g^2}{m_N}+\frac{g^2}{m_N-\epsilon_3}
\frac{2\epsilon_2}{m^2_\sigma-m^2_\pi}\nonumber\\
&\simeq  \frac{\epsilon_3 +2\epsilon_2 m_N/m^2_\sigma}{f^2_\pi} \simeq\frac{\Sigma (0)-m_Nm^2_\pi/m^2_\sigma}{f^2_\pi}
\,.\label{DoffAdler}
\end{eqnarray}
Taking  $t=0$ in (\ref{Doff}), in the Weinberg point we have
\begin{eqnarray}
&\widetilde{D}^+(0,0,0,0)=
\frac{g^2}{m_N-\epsilon_3}-\frac{g^2}{m_N}+\frac{g^2}{m_N-\epsilon_3}
\frac{2\epsilon_2-m^2_\pi}{m^2_\sigma}\nonumber\\
&\simeq -\frac{m_N m^2_\pi }{m^2_\sigma f^2_\pi}+ \frac{\epsilon_3 +2\epsilon_2 m_N/m^2_\sigma}{f^2_\pi}\simeq\frac{\Sigma (0)-2m_Nm^2_\pi/m^2_\sigma}{f^2_\pi}
\,.\label{DoffWeingerg}
\end{eqnarray}

Note that  both the axial current divergence has its canonical form (\ref{PCAC2}) and the conditions (\ref{Weinberg}), (\ref{Adler}),  (\ref{ChengDashen}), as well as (\ref{A2}),  are fulfilled at $\epsilon_2=\epsilon_3=0$, $\epsilon_1 =f_\pi m^2_\pi$.
However, as we see, the    conditions (\ref{Weinberg}), (\ref{Adler}),  (\ref{ChengDashen}),  as well as (\ref{A2}),  cf.  (\ref{Doffapp}),  (\ref{ChD}), (\ref{DoffAdler}), (\ref{DoffWeingerg}),  are also satisfied  at a weaker assumption that (\ref{cond}) is valid and in the latter case the axial current divergence also gets its canonical form (\ref{PCAC2}),  provided one may retain only first term in (\ref{langl}) and put $\langle \sigma\rangle =f_\pi$, i.e., if the tree approximation is valid.

\subsection{S-wave pion polarization operator and mean pion field}

Closing the nucleon legs in diagrams for the  amplitude of the forward $\pi N$ scattering shown in Fig. \ref{PiNAmpl} (evaluated in the tree approximation
to the $\pi N$ scattering) one finds the pion polarization operator in the gas approximation, cf. \cite{Migdal:1978az,Migdal:1990vm}.
The non-pole part of the pion polarization operator in the isospin-symmetric matter  consists of the s-wave and p-wave parts,
\begin{eqnarray}
&\Pi_S (\omega, \vec{q}, N=Z)+\delta \Pi_P =-\widetilde{D}(\nu,  q=q^{\prime})n\label{Ps}\\
&+Bn^{1+\alpha}\,,\nonumber
\end{eqnarray}
where $n=n_p +n_n$ is the nucleon density, $\alpha >0$.  Here  $\delta \Pi_P(\omega, \vec{q})\propto \vec{q}\vec{q}^{\,\prime} n$ is taken at $\vec{q}=\vec{q}^{\,\prime}$. It should be added to the pole $NN^{-1}$  contributions to the p-wave pion polarization operator, $\Pi_P$, where $N^{-1}$ is the nucleon hole. $\Delta (1232)$ isobars can be also included, cf. \cite{Baym:1975tm,Migdal:1990vm}. In the gas approximation   terms $\propto Bn^{1+\alpha}$, being higher-order in $n$,  should be dropped. We further assume that for all densities of our interest here the correlation effects $\propto B$  remain to be   small, i.e., that  $|Bn^\alpha|\ll |\widetilde{D}|$. A rough estimate  \cite{Kolomeitsev:2002gc} gives
$\alpha =1/3$, $B\sim 1/(\pi^2 m^2_\pi)$.

The  effective pion Lagrangian density of the  classical one-Fourier-component charged pion field $\phi_{\omega,\vec{q}}$ is as follows \cite{Migdal:1978az,Migdal:1990vm},
\begin{eqnarray}
L_{\rm ef}(\omega, \vec{q}, n_n, n_p, \phi_{{q}})=\langle L(\omega, \vec{q})\rangle_N\,,\nonumber
\end{eqnarray}
where averaging is done over the nucleon medium and we retained the dependence on the classical pion field up to second-order terms. We have
\begin{eqnarray}
L_{\rm ef}(\omega, \vec{q}=\vec{q}^{\,\prime})=(\omega^2-\vec{q}^{\,2} -\mbox{Re}\Pi (\omega, \vec{q},n ))|\phi_{\omega,\vec{q}}|^2\,,\label{efL}
\end{eqnarray}
where  $\Pi \simeq \Pi_P (\omega, \vec{q}\vec{q}^{\,\prime})|_{\vec{q}=\vec{q}^{\,\prime}}+\delta \Pi_P(\vec{q}\vec{q}^{\,\prime})|_{\vec{q}=\vec{q}^{\,\prime}} + \Pi_S$, i.e. it includes the p-wave and the s-wave terms. Certainly such a calculated polarization operator includes  not all possible diagrams. The resummation is done following the Dyson equation $D=D_0+D_0\Pi D$, where $D_0$ is the free pion Green function with $\Pi_P$ which phenomenologically includes the $NN$ correlations in the vertices, cf. \cite{Migdal:1978az,Migdal:1990vm}, and  with the term $\Pi_S+\delta \Pi_P$, as we evaluated it  in the gas approximation.  The nucleons  are treated, as quasiparticles with the effective Fermi liquid nucleon mass, $n$ is  the full baryon density.

The spectrum of quasiparticle excitations with the pion quantum numbers is found from (\ref{efL}) when one puts \begin{eqnarray}
D^{-1}(\omega, \vec{q},n)=\omega^2 -\vec{q}^{\,2}-\Pi (\omega, \vec{q},n)=0\,.\label{disp}
\end{eqnarray}

\subsection{SM1 model}

Assuming  that (\ref{cond}) is approximately fulfilled, we may following (\ref{sigmat})
  put $\Sigma (t=2m^2_\pi)\simeq \Sigma (t=0)=\Sigma_1 =m_N m^2_\pi/m^2_\sigma$. From (\ref{Doffapp}), (\ref{cond}) we have
\begin{eqnarray}
&\widetilde{D}_{\rm SM1}^+ (\nu, t,q^2, q^{\prime\,2})\simeq \frac{m_N (t-m^2_\pi)}{m^2_\sigma f^2_\pi}\nonumber\\
&\simeq \frac{\Sigma_1 (q^2+q^{\prime\,2}-m^2_\pi -2\omega^2)}{f^2_\pi m^2_\pi}+\frac{2\Sigma_1}{f^2_\pi m^2_\pi}\vec{q}\vec{q}^{\,\prime}
\,,\label{Dsigmt}
\end{eqnarray}
where we label the amplitude by subscript 1 provided condition (\ref{cond}) is supposed to be fulfilled. Putting $t=0$, $q^2=q^{\prime\,2}=m^2_\pi$,  we find the relation between the $\pi N$ scattering length and the $\Sigma$-term in this model:
$4\pi a^{+}_{\pi N} (1+m_\pi/m_N)=-\Sigma_1/f^2_\pi$. However this relation  badly agrees with the experimental value  $a^+_{\pi N}\simeq -0.0083/m_\pi$, which is a tiny quantity. Taking (\ref{Dsigmt}) on the mass-shell, i.e. for $q^2=q^{\prime\,2}=m^2_\pi$, one would get
\begin{eqnarray}
&\widetilde{D}_{\rm SM1}^+ (\nu, t(q^2=q^{\prime\,2}=m^2_\pi))\nonumber\\
&\simeq \frac{\Sigma_1 (m^2_\pi -2\omega^2)}{f^2_\pi m^2_\pi} +\frac{2\Sigma_1\vec{q}\vec{q}^{\,\prime}}{f^2_\pi m^2_\pi}
\,,\label{Dsigmt1}
\end{eqnarray}
again being in contradiction with the condition that $a^+_{\pi N}$ is a tiny quantity.

Using  (\ref{Dsigmt}) in the gas approximation (for $B=0$) we find
\begin{eqnarray}
\Pi_{S}^{\rm SM1, off} (\omega, \vec{q}, N=Z)\simeq \Sigma_1 (m^2_\pi +2\vec{q}^{\,2})n/(m^2_\pi f^2_\pi)\,,\label{Pilsm}
\end{eqnarray}
and
\begin{eqnarray}
&\delta\Pi_{P}^{\rm SM1} (N=Z)\simeq -2\Sigma_1 \vec{q}\vec{q}^{\,\prime}n/(m^2_\pi f^2_\pi)\,\label{deltaP}
\end{eqnarray}
for $\vec{q}=\vec{q}^{\,\prime}$.
Summing (\ref{Pilsm}), (\ref{deltaP}) we  have
\be\Pi_{S}^{\rm SM1, off}+\delta\Pi_{P}^{\rm SM1} (N=Z)\simeq \Sigma_1 n/ f^2_\pi\,.\label{PiSPof}\ee
Superscript ``off'' indicates that  $\Pi_S^{\rm SM, off}$ is constructed with the help of the amplitude
taken at arbitrary values $\omega$ and $\vec{q}$.
Thus in this case (with $\Pi_{S}^{\rm SM1, off}$) the s-wave $\pi N$ interaction  in the isospin-symmetric matter derived within the linear sigma model proves to be repulsive and the s-wave pion condensation at $\omega_c  =0$ does not occur
at least up to sufficiently  high densities. The term $\Sigma_1 n/ f^2_\pi$ can be included with the help of the  replacement $m^2_\pi \to m^2_\pi +\Sigma_1 n/ f^2_\pi$.
After this replacement is done, the consideration of the problem of the p-wave pion-nucleon condensation   performed in \cite{Migdal:1971cu,Migdal:1972,Scalapino:1972fu,Sawyer:1972cq,Migdal:1973bwk,
  Migdal:1973PL,Migdal:1973PL,Migdal:1973jkf,Migdal:1973zm,Baym:1973zk,Migdal:1974jn,
  Campbell:1974qt,Campbell:1974qu,Baym:1975tm} and other works, cf. reviews \cite{Brown:1975di,Migdal:1978az,Voskresensky:1982vd,Ericson:1988gk,Migdal:1990vm},  remains the same.

The pion spectrum  in isospin-symmetric matter calculated using Eqs. (\ref{Pilsm}), (\ref{deltaP}) (in model ``SM1,off'')  renders
  \begin{eqnarray}
\omega^2_{\rm SM1, off} =m^2_\pi + \frac{\Sigma_1 n}{f^2_\pi}  +\vec{q}^{\,2}
+ \Pi_P^{\rm SM}(\omega, \vec{q},N=Z)
\,.\label{spectrumSM}
\end{eqnarray}

\subsection{SM2 model}

Now do not require fulfilment of the condition (\ref{cond}). Then we can satisfy condition  $a^{+}_{\pi N}\simeq 0$, but  at the price of the choosing of a large value for the $\Sigma$ term in (\ref{sigmat}), $\Sigma =\Sigma_2\simeq 2m_Nm^2_\pi/m^2_\sigma$, in this case we label the amplitude by subscript 2. Assuming, as above,  $m_\sigma\simeq 600$ MeV we get  $\Sigma_2\simeq 100$ MeV.
Then from (\ref{sigmat}), (\ref{Doffapp}) we  obtain
\begin{eqnarray}
\widetilde{D}_{\rm SM2}^+ (\nu, t,q^2, q^{\prime\,2}, a^{+}_{\pi N}=0) =m_N t/(m^2_\sigma f^2_\pi)\,.\label{Da}
\end{eqnarray}
Weinberg and Adler conditions (\ref{Weinberg}) and (\ref{Adler}), as well as (\ref{A2}), prove to be violated.
However note that even in this case the amplitude $\widetilde{D}_{\rm SM2}^+ $  as a function of $t$ shows a smooth change from the Cheng-Dashen point through the Adler point, where now $\widetilde{D}_{\rm SM2}^+ (0,m^2_\pi,m^2_\pi,0)=\Sigma_2/2f^2_\pi$,  to the Weinberg point where now $\widetilde{D}_{\rm SM2}^+ (0,0,0,0)=0$. The corresponding on-mass-shell limit of the amplitude is given by
 \begin{eqnarray}
 &\widetilde{D}_{\rm SM2}^+ (\nu, t(q^2=q^{\prime\,2}=m^2_\pi), a^{+}_{\pi N}=0)\nonumber\\
 &=2m_N (m^2_\pi-\omega^2+\vec{q}\vec{q}^{\,\prime})/(m^2_\sigma f^2_\pi)\,.\label{Daon}
\end{eqnarray}
Thus within the sigma model choosing $m_\sigma \simeq 600$ MeV  we may either satisfy the  experimental finding that $\Sigma \simeq (50-60)$MeV  violating requirement that $a^+_{\pi N}\simeq 0$ or we may fulfill condition $a^+_{\pi N}\simeq 0$ at the price of the usage of an increased value of $\Sigma\simeq 100$ MeV and at  violation of the Weinberg and Adler conditions  (\ref{Weinberg}), (\ref{Adler}). Note  that  taking  a  larger value for  $m_\sigma$ we could, at this price, decrease the quantity $\Sigma$ to the values not contradicting the data.

In case of the model  2 we have
\be\Pi_{S}^{\rm SM2, off}+\delta \Pi_{P}^{\rm SM2, off}=0\,,\label{PSP2of}\ee
as it follows from (\ref{Da}) for $q=q^{\prime}$, i.e.  $t=0$, and as it has been used in the early works \cite{Migdal:1974jn,Campbell:1974qt,Campbell:1974qu,Baym:1975tm}.  As in case of model 1, the consideration of the possibility of the p-wave pion condensation performed in \cite{Migdal:1971cu,Migdal:1972,Scalapino:1972fu,Sawyer:1972cq,Migdal:1973bwk,
  Migdal:1973PL,Migdal:1973PL,Migdal:1973jkf,Migdal:1973zm,Baym:1973zk,Migdal:1974jn,
  Campbell:1974qt,Campbell:1974qu,Baym:1975tm} and other works, cf. reviews \cite{Brown:1975di,Migdal:1978az,Voskresensky:1982vd,Ericson:1988gk,Migdal:1990vm},  remains unchanged.

Employing the amplitude (\ref{Da}) (model ``SM2,off'') we  arrive at the  spectrum
  \begin{eqnarray}
  \omega^2_{\rm SM2, off} =m^2_\pi +\vec{q}^{\,2}+\Pi_P^{\rm SM}(\omega, \vec{q},N=Z)\,.
\label{spectrumSM2of}
\end{eqnarray}
Namely such a spectrum (taken at  $\Pi=\Pi_P$) has been studied in  \cite{Campbell:1974qt,Campbell:1974qu,Baym:1975tm,Voskresensky:1982vd} in case of the isospin-symmetric matter.

\subsection{Off-shell and on-shell treatments of problem of  s-wave pion condensation}

In case when  the Lagrangian density is fixed, here by Eqs. (\ref{linsigm}), (\ref{linsigm1}), (\ref{linsigm0}),
the
off-mass-shell amplitude has certain  physical sense,  it  permits  to construct the pion polarization operator in the nucleon medium in the gas approximation and the effective pion Lagrangian, which  allow to describe specific observable effects, e.g., such as the possibility of the critical phenomena associated with  the pion condensate phase transition. The same pion polarization operator follows in the gas approximation, if we use the ordinary Feynman diagrammatic rules.
This  is the key statement for our further consideration. It
was put in doubt in  \cite{Delorme92,Ericson94} and in a number of  subsequent works. Constructing $\Pi_S$,  those authors supposed not only to put $q=q^{\prime}$ but also take  $q^2=q^{\prime\,2}=m^2_\pi$  with the argument that  the scattering amplitude for particles in vacuum has the meaning   only for on-mass-shell particles, i.e. for $q^2=q^{\prime\,2}=m^2_\pi$.
In reality, pions can be considered as free particles only between collisions with  rarely distributed infinitely massive centers,  when the nucleon recoil and pion coherence effects can be neglected.
The statement should not work for consideration of the classical pion field in a rather dense matter.
Anyhow in the case of the sigma model under consideration with such a on-mass-shell approach, using the amplitude (\ref{Dsigmt1}) one would arrive at
\begin{eqnarray}
\Pi_{S}^{\rm SM1, on} (\omega, \vec{q}, N=Z)
\simeq \frac{\Sigma_1 (2\omega^2-m^2_\pi)n}{f^2_\pi m^2_\pi}\label{on}
\end{eqnarray}
 with $\Sigma_1 =m_N m^2_\pi/f^2_\pi$,
  and  one would get
  \begin{eqnarray}
\omega^2_{\rm SM1, on} =\frac{m^2_\pi (1-\frac{\Sigma_1  n}{f^2_\pi}m^2_\pi) + \Pi_P^{\rm SM}
}{1-2\Sigma_1 n/(f^2_\pi m^2_\pi)}+ \vec{q}^{\,2}\,,\label{spectrumSM1}
\end{eqnarray}
that differs from (\ref{spectrumSM}), although only in terms obtained beyond the framework of the validity of the gas approximation.

 With (\ref{Daon})  one would obtain
\begin{eqnarray}
\Pi_{S}^{\rm SM2, on} (\omega, \vec{q}, N=Z)
\simeq \frac{\Sigma_2 (\omega^2-m^2_\pi -\vec{q}\vec{q}^{\,\prime})n}{f^2_\pi m^2_\pi}\,\label{Pon}
\end{eqnarray}
 at $\Sigma_2 =2m_N m^2_\pi/f^2_\pi$ and $\vec{q}=\vec{q}^{\,\prime}$ and
  \begin{eqnarray}\omega^2_{\rm SM2, on} =m^2_\pi+\frac{  \Pi_P^{\rm SM}
  }{1-n\Sigma_2/(f^2_\pi m^2_\pi)}+\vec{q}^{\,2}\,\label{spectrumSM2}
\end{eqnarray}
instead of (\ref{spectrumSM2of}). We see that the squared effective pion mass determined, as the quantity entering the spectrum $\omega^2 (\vec{q}=0)=m^{*\,2}$,  in the approximation linear in $n$, is the same  for the models ``SM1,off''  and ``SM1,on'' ($m^{*\,2}_1 = m^2_\pi +\Sigma_1 n/f^2_\pi$)), and  for the models ``SM2,off'' and ``SM2,on'',
$m^{*\,2}_2=m^2_\pi \neq m^{*\,2}_1$.  However in nonlinear in $n$ terms the quantities $m^{*\,2}_{\rm {1,off}}$ and $m^{*\,2}_{\rm{1,on}}$ are different, as well as $m^{*\,2}_{\rm{2,off}}$ and $m^{*\,2}_{\rm{2,on}}$. This result does not disagree with the equivalence theorem, cf. \cite{Kondratyk2003}, since the conditions of applicability of the equivalence theorem are not fulfilled in this case.

 Note that Ref. \cite{Brown91} associated occurrence of the s-wave pion condensation with the vanishing of the effective pion mass term in the effective Lagrangian, however in their model $\omega$ dependence  in $\Pi_S$ was not included.
  In case of model ``SM1,on''  the effective pion mass defined as $m^2_{\rm ef}=m^2_\pi +\Pi (q^2=0)$ vanishes for
$n> n_{cm}^{\rm SM1,on}=  f^2_\pi m^2_\pi/\Sigma_1 \simeq (2-2.5) n_0$.    Assuming decrease of the effective pion decay parameter  $f_\pi$ with increasing density Ref. \cite{Brown91} estimated a still smaller value  $n_{cm}\simeq 1.6 n_0$. In case of model ``SM2,on'' such a defined the effective pion mass vanishes for
$n> n_{cm}^{\rm SM2,on}=  f^2_\pi m^2_\pi/\Sigma_2$. However then Refs. \cite{Delorme92,AdamiBrown93} found that a repulsion from the range $\omega^2$ term compensates the attraction in the  expression for the pion spectrum.
In  case ``SM1,on''  resulting in   the spectrum (\ref{spectrumSM1}), the effective pion mass $m^*$ ($\neq m_{\rm ef}$) increases with increasing $n$ and in  case ``SM2,on''  resulting in   the spectrum (\ref{spectrumSM2}), $m^*$ stays constant.
Also, we see that expressions for spectra contain  poles: the spectrum given by Eq. (\ref{spectrumSM1})   at
$n= n_{c\omega}^{\rm SM1,on}=  f^2_\pi m^2_\pi/(2\Sigma_1)$, and the spectrum given by  Eq. (\ref{spectrumSM2}) for
$n= n_{c\omega}^{\rm SM2,on}=  f^2_\pi m^2_\pi/\Sigma_2$.
With $\Sigma_2\simeq 2\Sigma_1 \simeq 100$ MeV we estimate
$n_{c\omega}^{\rm SM1,on}=n_{cm}^{\rm SM1,on}/2$, $n_{c\omega}^{\rm SM2,on}=n_{cm}^{\rm SM2,on}\simeq n_{cm}^{\rm SM1,on}/2$, and thus 
  $n_{c\omega}^{\rm SM1,on}\simeq n_{c\omega}^{\rm SM2,on}\simeq (1-1.3) n_0$ in both models.
References \cite{FriedmanGal20,FriedmanGal2007}   employed    $f_\pi (n)$ instead of $f_\pi$ in the description of pion atoms. With $f_\pi \to f_\pi (n)$ the value $n_{c\omega}$ is still decreased. However note that presence of the pion condensation at $n<n_0$ contradicts to the experimental data.

The possibility of the s-wave pion condensation in isospin-symmetric matter was  not worked out  in \cite{AdamiBrown93,Delorme92} and in subsequent papers,  which  employed the s-wave polarization operator found with the help of the on-mass-shell scattering amplitude in their models.  Most of the researches focused  attention on  the s-wave kaon condensation in isospin-asymmetric matter, although consideration of the kaon condensation problem is  completely analogous to the consideration of the pion condensation. The point is that Ref. \cite{Delorme92,AdamiBrown93}  associated possibility of the s-wave pion condensation with  vanishing of the squared effective pion mass in the pion spectrum.
In their on-shell model, for $a^+_{\pi N}$, being small negative quantity, the value   $n_{cm}>n_{c\omega}$.
The authors \cite{Delorme92} wrote that, when  the overall coefficient of the terms  $\propto\omega^2$ in the expression for the spectrum given by $D^{-1}=0$
vanishes at $n=n_{c\omega}$, ``it forces the squared wave number $\vec{q}^{\,2}$ to be negative for any $\omega$. This situation corresponds to evanescent waves in the medium,'' as they stated, rather than to the pion condensation.  For $n>n_{cm}>n_{c\omega}$ the quantity $m^{*\,2}$ is positive in their model, cf. Eq. (26) in \cite{Delorme92}, not allowing for the s-wave condensation according their argumentation.
We do not support this statement of \cite{Delorme92}.
In the models ``SM1,on'' and ``SM2,on'' the term proportional to the squared effective pion mass, $m_{\rm ef}^2$,
 in the effective pion Lagrangian density,  $\delta L^{\rm SM,on}=m^2_\pi |\phi|^2(n/n_{cm}^{\rm SM,on} -1)/2$,  changes the sign  for $n>n_{cm}^{\rm SM,on}$.
 This means  that the s-wave pion condensation associated with the change of the sign of the squared effective pion mass term, $m_{\rm ef}^2$, in the energy density is energetically favorable
 at $n>n_{cm}^{\rm SM,on}$ in the given models. 
 Also note that  in both  sigma models, ``SM1,on'' and ``SM2,on'', the energy density acquires the $\omega^2$-dependent term $\delta E^{\rm SM,on}_\omega =-\omega^2 |\phi|^2(n/n_{c\omega}^{\rm SM,on} -1)/2$ that may result in appearance of a frequency dependent classical pion field  for $n>n^{\rm SM,on}_{c\omega}$.  We continue discussion of the s-wave pion condensation  in Sect. \ref{Tobe}.

 We should stress that Eqs. (\ref{Pilsm}) and (\ref{deltaP}) for $\Sigma_1 =m_N m^2_\pi/f^2_\pi$ as well as  Eq. (\ref{PSP2of})   for $\Sigma_2 =2m_N m^2_\pi/f^2_\pi$   follow in the low $n$ approximation right from the Lagrangian of the model  (in the tree approximation for the amplitude) and  do not need extra putting $q^2=q^{\prime\,2}=m^2_\pi$. To get them we just closed the nucleon legs in the diagrams shown in Fig. \ref{PiNAmpl} that corresponds to the averaging of the Lagrangian over the nucleon degrees of freedom. Although for a small $n$ the probability of the multiple scattering processes is also small,   it is principally nonzero, and its calculation  requires integrations over $\omega$ and $\vec{q}$ rather than putting $q^2=m^2_\pi$.  Moreover we should say that the gas approximation in calculation of $\Pi_S$ might  be  is valid up to   $n\lsim {\rm several} \,\,n_0$, because the density dependence of the diagrams with multiple integrations in the intermediate states entering $\Pi_S$ is a weak.  Thereby, Refs. \cite{Migdal:1978az,Migdal:1990vm,Voskresensky:1993ud}   treated expression (\ref{Ps}) with $B=0$, as approximately valid not only for $n\ll n_0$ but  in a wider range of densities.  Moreover, a numerical smallness of some specific diagrams contributing to $\Pi_S$ beyond the validity of the gas approximation was demonstrated in \cite{Kolomeitsev:2002gc,Kolomeitsev:2002mm}. Also, a successful fit of the phenomenological pion optical potential to the pion atom data \cite{Troitsky:1981,Kolomeitsev:2002gc,FriedmanGal20,FriedmanGal2007}  was performed employing  smallness of the correlation effects in $\Pi_S$. Thus following our argumentation,  $\omega$ and $\vec{q}$  in (\ref{Ps}) should be treated as  independent variables, not connected by the on-mass-shell condition.

Some works followed the idea of   Ref.  \cite{BrownLeeRho} that the answer on the question about fulfilment or not fulfilment of the PCAC condition (\ref{PCAC}) and the current algebra theorems (\ref{Weinberg}), (\ref{Adler}),  (\ref{ChengDashen}) depends  on   the choice of the artificial interpolating fields, in which terms one may rewrite  the  Lagrangian, not changing  physics. We should stress that Eqs. (F4), (F5) of \cite{BrownLeeRho}, which they use to demonstrate their point,  do   depend on the choice of the interpolating fields,  since $\delta L=j_A\phi \neq j_A\widetilde{\phi}$. Thereby the Lagrange equations, which follow from such constructed  Lagrangians, are different reflecting difference in   physical effects associated with presence of even tiny $j_A\neq 0$. Thus the statement on relevance of the off-shell effects in the problem of the s-wave pion condensation does not contradict to often mentioned equivalence theorem that  any  local change of variables in quantum field theories, which leaves the free field part of the Lagrangian unchanged, does not alter the $S$-matrix \cite{Chisholm1961}, and it does not contradict to on-shell
consideration of the  scattering of particles on infinitely heavy centers \cite{AgassiGal}.
However, already with the  two-particle scattering in vacuum obeying the Bethe-Salpeter
equation in the tree-level kernel there appear complications, since solutions in this kernel  prove to be dependent on the representation of the theory \cite{Kondratyuk2001}. In our case nucleons in matter undergo recoil effects  and the pion field (especially the classical condensate field) does not obey the free Klein-Gordon equation.

Let us also mention that similar results, as for the ``SM2,off'', follow from  an extended linear sigma model \cite{Myhrer1}, permitting to reproduce the experimental value of the axial-vector constant $g_A$. Essentially increasing the parameter $m_\sigma$  in the ordinary and extended sigma models it is  possible to recover  values both of the $\pi N$ scattering length and the sigma-term.  However one should notice that the value  $m_\sigma \sim 600$ MeV is required to get an appropriate fit of the nuclear equation of state in the relativistic mean-field models, cf. \cite{Kolomeitsev:2004ff}. Also, both the linear sigma model and the extended linear sigma model hardly reproduce  some of
the $\pi N$ scattering amplitude properties
predicted by the heavy-baryon chiral perturbation theory, cf.  \cite{Myhrer2}.

Concluding this section, we argued that with the once chosen  Lagrangian of the model (here  the sigma model),   at the lowest tree level order, in the low density approximation, the pion polarization operator is unambiguously constructed. The s-wave part of the polarization operator  is proportional to the pion-nucleon scattering amplitude at  arbitrary relation between kinematical variables,
rather than to the on-mass-shell amplitude  $\widetilde{D} (\nu,t,q^2=q^{\prime\,2}=m^2_\pi)$ taken at $\vec{q}=\vec{q}^{\,\prime}$. Usage of the on-mass-shell amplitude to construct $\Pi_S^{\rm SM}$  within the models  SM1 and SM2
in the gas approximation over nucleons and for the classical pion mode, may lead  to erroneous results, such as occurrence of the s-wave pion condensation in the  isospin-symmetric matter at rather low densities. However in application to other models the issue can be   more subtle.

\section{Model of Manohar-Georgi}\label{MonaharGeorgi}

To continue demonstration of our key point that the knoweledge of the scattering amplitude off-mass-shell matters in the problem of the s-wave pion polarization in the medium let us consider  two  other Lagrangians, introduced in \cite{ManoharGeorgi84} and \cite{GasserSainioSvarc88}.
With a reduction to
$SU(2)_L \times SU(2)_R$ and the s-wave channel, the $O(Q^2)$ Manohar-Georgi (MG) Lagrangian density renders
\begin{eqnarray}
&L_{\rm MG}=\bar{N}iv^\mu \partial_\mu N- \Sigma  \bar{N}N
+\partial_\mu \vec{\pi}\partial^\mu \vec{\pi}/2-m^2_\pi \vec{\pi}^2/2+L_{\rm WT}\nonumber\\
&+[\Sigma \vec{\pi}^2/2 +c_2 (v^\mu \partial_\mu\vec{\pi})^2+c_3 \partial_\mu \vec{\pi}\partial^\mu \vec{\pi}]\bar{N}N/f^2_\pi...\,,\label{MG}
\end{eqnarray}
cf.
\cite{ThorssonWirsba95}, where $v^\mu$ is the 4-velocity of the nucleon and $v^\mu = (1,0,0,0)$ in the center-of-mass reference frame of the nucleon, the values $m^2_\pi$ and $f^2_\pi$ include $O(Q^2)$ loop corrections to
the corresponding quantities at tree level, constants $c_2$ and $c_3$ are of the order
$O(Q^0)$ and constant $\Sigma =\Sigma(0)$ is linear in the quark masses and
therefore is of order  $O(Q^2)$. The Weinberg-Tomozawa (vector) term, $\Pi_S^{\rm WT,-} \simeq (n_n-n_p) \omega/(2f^2_\pi)$, does not enter into
the pion self-energy in the case of the isospin-symmetric nuclear matter of our interest here. We should note that the original MG Lagrangian contains infinite number of  terms (labeled by dots in (\ref{MG})), whereas we shall consider the reduced MG Lagrangian given by Eq. (\ref{MG}), i.e. dropping   terms labeled by dots.

The Lagrangian density (\ref{MG}) yields the amplitude
\begin{eqnarray}
\widetilde{D}^+_{\rm MG} \simeq \frac{2c_2\omega\omega^{\prime}+2c_3 qq^{\prime}+\Sigma}{f^2_\pi}\,,\label{DMGappr}
\end{eqnarray}
i.e., $\widetilde{D}^+ \simeq \Sigma/f^2_\pi$ in all three kinematical points (\ref{Weinberg}), (\ref{Adler}),  (\ref{ChengDashen})
at $\vec{q}=0$, and thereby  conditions (\ref{Weinberg}), (\ref{Adler}) are not fulfilled. Only the lowest order $\vec{\pi}^2 \bar{N}N$ terms were included to derive Eq. (\ref{DMGappr}).

 From the reduced Lagrangian density (\ref{MG}) one recovers the isospin-even pion-nucleon scattering length,
 \begin{eqnarray}
 a^+_{\pi N}\simeq \frac{(2c_2+2c_3)m^2_\pi +\Sigma}{4\pi f^2_\pi (1+m_\pi/m_N)}\,.
 \end{eqnarray}
  We find $(2c_2+2c_3)m^2_\pi \simeq -\Sigma +4\pi b_{+}$, $b_{+}=a^+_{\pi N} f^2_\pi (1+m_\pi/m_N)$.     The tiny value $4\pi b_{+}$ is further put zero, as we have done above within the sigma model.

 Then
\begin{eqnarray}
\widetilde{D}^+_{\rm MG} \simeq \frac{\Sigma (m^2_\pi -\omega^2)}{f^2_\pi m^2_\pi}  -\frac{2c_3 \vec{q} \vec{q}^{\,\prime}}{f^2_\pi}\,,\label{DMGpl}
\end{eqnarray}
and in the gas approximation
\begin{eqnarray}
\Pi_S^{\rm MG} (\omega, \vec{q}, N=Z)\simeq  n\Sigma \frac{\omega^2 -m^2_\pi }{f^2_\pi m^2_\pi}
\,,\label{PiMG}
\end{eqnarray}
and
\begin{eqnarray}
\delta \Pi_P^{\rm MG}\simeq  2c_3 \vec{q} \vec{q}^{\,\prime}n/f^2_\pi\,\label{PiPdMG}
\end{eqnarray}
is taken at $\vec{q} =\vec{q}^{\,\prime}$. These expressions remind expressions for
$\Pi_{S}^{\rm SM1,2,on}$, $\delta\Pi_P^{\rm SM}$ introduced in previous section. However here
expressions (\ref{PiMG}) and (\ref{PiPdMG}) follow right from the Lagrangian and one does not need to do additionally the replacement  $\vec{q}^2 =\vec{q}^{\,\prime\,2}=m^2_\pi$.

The spectrum of the pion quasiparticle excitations in the isosppin-symmetric matter
 is determined by putting zero the inverse pion  quasiparticle propagator (\ref{disp}). Thus from (\ref{PiMG}), (\ref{disp})  we get
\begin{eqnarray}
\omega^2 =m^2_\pi +\frac{\vec{q}^{\,2}(1+\frac{2c_3 n}{f^2_\pi}) + \Pi_P}{1-n\Sigma/(f^2_\pi m^2_\pi)}\,.\label{spectrumMG}
\end{eqnarray}

\begin{figure}\centering
\includegraphics[width=8.0cm,clip]{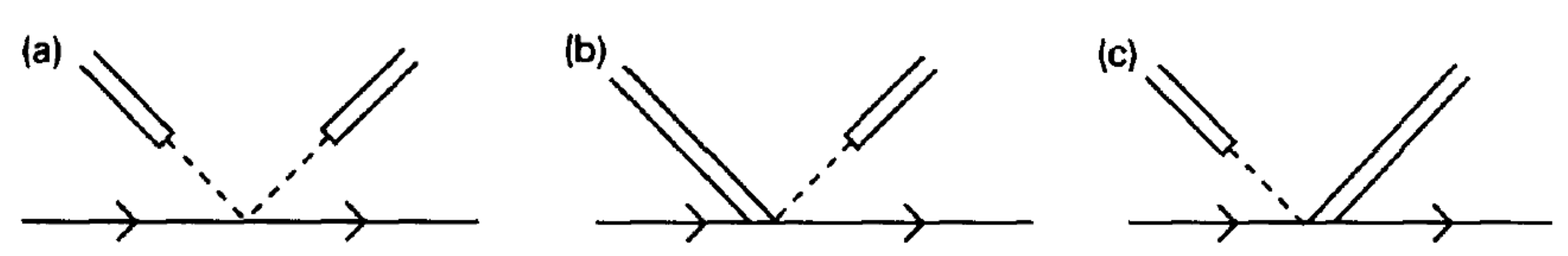}
\caption{Feynman diagrams contributing to $\pi N$ scattering amplitude in tree approximation  to MG model (a), and  to GSS model, (a), (b), (c). Solid line indicates nucleon, double-line relates to the pseudoscalar source $j_i$ and dashed line,  to pion.
 }\label{GSSfig1}
\end{figure}

Note that the same amplitude can be found from the Lagrangian density
\begin{eqnarray}
L_{\rm MG}\to L_{\rm MG}+j_i\pi_i \label{MGsource}
\end{eqnarray}
 at the  pseudoscalar source term considered in the limit  $j_i\to 0$. Then the amplitude is determined by the diagram (a) shown in Fig. \ref{GSSfig1} after amputation of the external legs. The point-vertex in Fig. \ref{GSSfig1} (a) is given by $i\widetilde{D}^+_{\rm MG}$.

Finally note that the MG model allows for the s-wave pion condensation in the isospin-symmetric matter for $n>n_c=f^2_\pi m^2_\pi /\Sigma$ provided the gas approximation holds up to such densities, cf.  discussion of models ``SM1,on'' and ``SM2,on'' in previous section  and consideration  below in Sect. \ref{Tobe}.

\section{Model of Gasser-Sainio-Svarc}\label{Gasser}

Within the functional integral formulation of chiral perturbation theory developed by Gasser
and Leutwiller \cite{GasserLeutwiller84}, which was extended to include nucleons, Gasser, Sainio and
Svarc  \cite{GasserSainioSvarc88} introduced the Lagrangian density (GSS),
\begin{eqnarray}
L_{\rm GSS}=L_{\rm MG}+j_i\pi_i (1-\Sigma \bar{N}N/f^2_\pi m^2_\pi)\,,\label{GSS}
\end{eqnarray}
with the pseudovector source $j_i=2Bf_\pi p_i$, $B=m^2_\pi/(m_u+m_d)$ satisfying the canonical PCAC condition (\ref{PCAC}) and with $L_{\rm MG}$ from (\ref{MG}). Since Green functions are obtained by taking functional derivatives of the generating functional with respect to the source $j_i$, the nontrivial coupling of the source to the pion field, here in the form $j_i\pi_i (1-\Sigma \bar{N}N/f^2_\pi m^2_\pi)$, matters.
From the Ward identity one gets \cite{GasserSainioSvarc88},
\begin{eqnarray}
&\widetilde{D}^+_{\rm GSS}=\frac{2c_2\omega\omega^{\prime}+2c_3 qq^{\prime}+\Sigma}{f^2_\pi}+ \frac{(q^2+q^{\prime 2}-2m^2_\pi)\Sigma}{m^2_\pi
f^2_\pi}\,,\label{DGSSWirsba}
\end{eqnarray}
cf. \cite{Kirchbach96}.
The same expression follows from the diagrams shown in Fig.  \ref{GSSfig1} for the connected $\pi\pi NN$ Green function in the lowest order \cite{ThorssonWirsba95}:
\begin{eqnarray}
&\widetilde{D}^+_{\rm GSS}=i(q^{2}-m^2_\pi) A^{\rm GSS}_{\pi N}(q^2,q^{\prime\,2})
(q^{\prime\,2}-m^2_\pi)\,,\nonumber\\
&A^{\rm GSS}_{\pi N}=\frac{i}{q^{2}-m^2_\pi}\frac{i(2c_2\omega\omega^{\prime}+2c_3qq^{\prime}+\Sigma)}{f^2_\pi}\frac{i}{q^{\prime\,2}-m^2_\pi}
\nonumber\\
&-\frac{\Sigma}{f^2_\pi m^2_\pi}\left(\frac{i}{q^{\prime\,2}-m^2_\pi}+\frac{i}{q^{2}-m^2_\pi}\right)\,.
\end{eqnarray}
The first term  in the last equality relates to the diagram shown in Fig. \ref{GSSfig1} (a) and the second term is associated with the second term in the Lagrangian (\ref{GSS}) and with two diagrams (b) and (c).
The off-mass-shell scattering amplitudes
satisfy now all three conditions (\ref{Weinberg}), (\ref{Adler}),  (\ref{ChengDashen}) in difference with the MG case. The condition (\ref{A2}) is fulfilled for $c_3=0$. For on-mass-shell variables, $q^2=q^{\prime\,2}=m^2_\pi$, the amplitudes $\widetilde{D}^+_{\rm GSS}$ and $\widetilde{D}^+_{\rm MG}$ coincide.

We could obtain the same Eq. (\ref{DGSSWirsba}) in another way. First perform variable replacement $\pi\to\pi+\alpha \pi (\bar{N}N)$ in the reduced MG Lagrangian dropping terms $O(\alpha^2, (\bar{N}N)^2)$. We get the Lagrangian density
\begin{eqnarray}
&L_\alpha =L_{\rm MG}
+\alpha [(\partial_\mu \bar{N}\cdot N+\bar{N}\cdot\partial_\mu  N)(\partial^\mu \pi^* \cdot \pi +\pi^*\partial^\mu \pi)]
\nonumber\\&
+2\alpha[\bar{N}N\partial_\mu\pi^*\partial^\mu \pi -m^2_\pi \bar{N}N\cdot\pi^*\pi]\label{cutMG}\\
&+ O(\alpha^2, (\bar{N}N)^2, \pi^3)\,,\nonumber
\end{eqnarray}
where the linear in $\alpha$ term follows from the contribution of the free pion Lagrangian.
In the Fourier transformation the terms linear in $\alpha$ yield in the amplitude the contribution
\begin{eqnarray}
\alpha (q-q^{\prime})^2) +2 \alpha (qq^{\prime}-m^2_\pi)=\alpha (q^2 +q^{\prime\,2}-2m^2_\pi)\,.\nonumber
\end{eqnarray}
Putting
$\alpha =\Sigma/(m^2_\pi f^2_\pi)$ we recover $\widetilde{D}^+_{\rm GSS}$
 and $A^{\rm GSS}_{\pi N}$ which fulfill the Cheng-Dashen, Adler and Weinberg conditions. Note that only, if we included all the dropped terms, the two models, labeled $\alpha$ and MG, would lead to identical results. Only in the latter case we could drop $\propto \alpha$ contribution relying on the  equivalence theorem. Although the term $\propto \alpha$  is obviously non-zero off mass shell in the  Lagrangian cutted at the order  $O(\alpha^2, (\bar{N}N)^2, \pi^3)$, it should be  compensated by the higher order diagrams corresponding to many-particle scatterings in the calculation of the observables, but only in case, when all the dropped terms are included. If one treats the Lagrangian (\ref{cutMG}) as it is, i.e.,  dropping the terms  $O(\alpha^2, (\bar{N}N)^2, \pi^3)$, values of the observables calculated using (\ref{cutMG}) and  (\ref{GSS}) for any $\alpha$
 and using (\ref{MG})   differ.

Let us consider explicitly  example of the static spatially uniform classical charged pion field $\phi =\langle(\pi_1+i\pi_2)/\sqrt{2}\rangle$. Then $|\pi|^2$ terms in $L_\alpha$ yield $\delta L_\alpha =\delta L_{\rm MG}
-2\alpha m^2_\pi \bar{N}N|\phi|^2+ O(\alpha^2, (\bar{N}N)^2, \phi^3)$, where $\delta L_{\rm MG}=-(m^2_\pi -\Sigma \bar{N}N/f^2_\pi)|\phi|^2$. The squared effective mass term of the field $\phi$ is given by $m^{*\,2}=m^2_\pi -(\Sigma/f^2_\pi -2\alpha m^2_\pi)\bar{N}N$. For $\alpha >\Sigma/(2m^2_\pi f^2_\pi)$ the term $m^{*\,2}|\phi|^2$ even changes the sign. Also we can see that with  the recovered  $\alpha^2$ contribution the partial term  $m^2_\pi|\phi^2|$ after the variable replacement would yield $ m^2_\pi|\phi|^2(1+\alpha \bar{N}N)^2$, whereas in the linear approximation we have $m^2_\pi|\phi^2|\to m^2_\pi|\phi|^2(1+2\alpha \bar{N}N)$. The former term is always non-negative, whereas the latter one changes the sign for $\alpha <-\Sigma/(2m^2_\pi f^2_\pi)$. By these examples we showed that  it is completely not surprising that   the reduced Lagrangians
(\ref{MG}) and (\ref{cutMG}) describe different physics.

Obviously the pion polarization operator, which is recovered from  the  reduced Lagrangian (\ref{cutMG}) with the help of the Feynman rule diagrammatics, also differs from that follows from the reduced MG Lagrangian. Note also that in the uniform gas approximation the term $\propto \alpha$ in the first line (\ref{cutMG}) is reduced to the full derivative in the effective action and can be dropped. However replacing  $t=(q-q^{\prime})^2
=2m^2_\pi -2qq^{\prime}$ following the on-mass-shell receipt we would get non-zero contribution  to the pion polarization operator in nuclear matter from this full derivative term.  This circumstance  can be considered as extra argument that the on-mass-shell replacement does not hold   for  calculation of the diagrams term by term and that the equivalence theorem  does not hold for Lagrangians (\ref{MG}) and (\ref{cutMG})
provided  $O(\alpha^2, (\bar{N}N)^2)$ terms are dropped.

The Lagrangian density $L_{\rm GSS}$ produces the s-wave part of the pion polarization operator  in the gas approximation
\begin{eqnarray}
\Pi_S^{\rm GSS} (\omega, \vec{q}, N=Z)\simeq  \frac{\Sigma n(m^2_\pi -\omega^2 +2\vec{q}^{\,2})
}{f^2_\pi m^2_\pi}
\,,\label{PiGSS}
\end{eqnarray}
where, as above, we used that $2(c_2+c_3) m^2_\pi \simeq -\Sigma$, and we have
$\delta \Pi_P^{\rm GSS}=\delta \Pi_P^{\rm MG}$.
Even for $\omega =m_\pi$ the terms  $\propto \vec{q}^{\,2}$ in $\Pi_S$ corresponding to  $L_{\rm MG}$ and $L_{\rm GSS}$ are different. Only for $q^2=m^2_\pi$, i.e. on mass shell, we get $\Pi_S^{\rm MG, on}=\Pi_S^{\rm GGS, on}$.

From (\ref{PiGSS}), (\ref{disp})  we obtain
\begin{eqnarray}
\omega^2 =m^2_\pi +\frac{\vec{q}^{\,2}(1+\frac{2\Sigma n}{f^2_\pi m^2_\pi}) + \Pi_P (\omega, \vec{q}\vec{q}^{\,\prime})+\delta\Pi_P }{1+n\Sigma/(f^2_\pi m^2_\pi)}\,,\label{spectrumGSS}
\end{eqnarray}
at $\vec{q}=\vec{q}^{\,\prime}$, that differs from (\ref{spectrumMG})  in the correlation terms.
It implies existence of  physical effects, which are different in the models described by the reduced Lagrangians $L_{\rm MG}$  and $L_{\rm GSS}$.
For example, as follows from (\ref{PiGSS}) at $\omega =0$,  likely the GSS model (with reduced Lagrangian in  off-mass shell treatment) does not allow for the s-wave pion condensation in isospin-symmetric matter, whereas the MG model allows it for $n>n_c\simeq (1.4-2.5)n_0$. However once more stress that  the diagrams in $\Pi_S$ beyond the gas approximation  in both cases were omitted.

\section{Phenomenological expressions for $\Pi_S$ in isospin-symmetric matter}\label{Phenomenological}
Now consider how one can proceed not employing  the microscopic expression for the  Lagrangian.
One may employ that the non-pole part of the amplitude $\widetilde{D}^+ (\nu,t,q^2,q^{\prime\,2})$  is a smooth
function of its variables. It can therefore be expanded
near the soft point ($q=q^{\prime}=0$) as a power
series in  $q^2$, $q^{\prime\,2}$, $\nu^2$ and  $t$.  In the static nucleon limit $\nu^2\simeq \omega^2$ and  $\nu_B =-qq^{\prime}/(2m_N)\to 0$. In this limit the pole term vanishes and $D=\widetilde{D}^+$.  Moreover one has $\nu_B =0$  in the Weinberg, Adler
and Cheng-Dashen kinematical  points, where thereby again $D=\widetilde{D}^+$.

Retaining only linear terms in the Taylor expansion of $\widetilde{D}^+ (\nu,t,q^2,q^{\prime\,2})$, after the regrouping the terms one arrives at
\begin{eqnarray}
&\widetilde{D}^+ (\nu,t,q^2,q^{\prime\,2})\simeq \alpha_1 +\alpha_2 (q^2+q^{\prime\,2})/m^2_\pi \nonumber\\
&+\beta\nu^2 +\gamma [(t-q^2-q^{\prime\,2})/2+\nu^2]+\beta_1\nu^4...,\label{DofDEE}
\end{eqnarray}
cf.  \cite{Delorme92}.
This expansion, although with differently regrouped terms, coincides with that previously employed in \cite{Migdal:1978az,Troitsky:1981,Migdal:1990vm}. For further needs we also explicitly wrote a higher order term $\beta_1\nu^4$.

As we have mentioned, the experimental value of $a^+_{\pi N}$ is a very small quantity. Thereby, simplifying consideration  we further continue to put  $a^+_{\pi N}\simeq 0$.
Estimated value $\beta_1\simeq 0.2/m_\pi$ is  small  \cite{Nagels1976,Migdal:1978az} (of the order of $\sim m_\pi/m_N$).
Dropping all terms $\sim m_\pi/m_N$ including the term $\propto \beta_1$,  we have
\begin{eqnarray}
&{D}^+ (m_\pi, 0,m^2_\pi,m^2_\pi)\simeq \widetilde{D}^+ (m_\pi, 0,m^2_\pi,m^2_\pi)\nonumber\\
&\simeq 4\pi (1+m_\pi/m_N)a^{+}_{\pi N}\simeq 0\,.\label{sct}
\end{eqnarray}
Additionally assuming fulfilment of  the Cheng-Dashen condition we have
\begin{eqnarray}
&\beta\simeq -(\Sigma -4\pi b_{+})/(f^2_\pi m^2_{\pi})\simeq  -\Sigma/(f^2_\pi m^2_{\pi})\,,\\
&\alpha_1+2\alpha_2 =\Sigma/f^2_\pi\,\nonumber
\end{eqnarray}
and we arrive at
\begin{eqnarray}
&\widetilde{D}^+ \simeq\alpha_1 +\frac{(\Sigma/f^2_\pi -\alpha_1)(q^2+q^{\prime 2})}{2m^2_\pi} -\frac{\Sigma \nu^2}{f^2_\pi m^2_\pi}\label{ChDMassSh}\\
&+\gamma \vec{q}\vec{q}^{\,\prime}\,.\nonumber
\end{eqnarray}
Employing  the Weinberg condition  we get  $\alpha_1 =-\Sigma/f^2_\pi$, $\alpha_2 =\Sigma/f^2_\pi$. The   Adler condition is then fulfilled automatically.
Using in (\ref{DofDEE}) that  $\nu^2\simeq \om^2$ and  $(t-q^2-q^{\prime\,2})/2+\nu^2\simeq \vec{q}\vec{q}^{\,\prime}$
 we arrive at
\begin{eqnarray}
&\widetilde{D}^+ (\omega^2,q^2,q^{\prime\,2},\vec{q}\vec{q}^{\,\prime})\simeq  -\frac{\Sigma (m^2_\pi -q^2-q^{\prime\,2}+\omega^2)}{f^2_\pi m^2_\pi}\label{DofDEE1} \\ &+\gamma \vec{q}\vec{q}^{\,\prime}\,.\nonumber
\end{eqnarray}
The  condition (\ref{A2}) is satisfied for $\gamma =\Sigma/(f^2_\pi m^2_\pi)$.
Recall that all three Weinberg, Cheng-Dashen and Adler conditions  are satisfied in the  GSS model described by the reduced Lagrangian $L_\alpha$ for $\alpha =\Sigma/(m^2_\pi f^2_\pi)$, and in the ``SM1,off'' model. The condition (\ref{A2}) is fulfilled in the ``SM1,off'' model, whereas in the GSS model described by the reduced Lagrangian $L_\alpha$ for $\alpha =\Sigma/(m^2_\pi f^2_\pi)$ it is satisfied for $c_3 =0$.
Note also that Eq. (\ref{DofDEE1}) can be considered as the simplest linear in $t$,  $q^2=q^{\prime\,2}$ and $\nu^2$ interpolation expression between the Cheng-Dashen and Weinberg points,  satisfying the mass-shell condition  (\ref{sct}).

With   the $\pi N$ non-pole amplitude (\ref{DofDEE1}), in the gas approximation we arrive at the s-wave pion polarization operator (labeled below as MSTV)
\begin{eqnarray}
&\Pi_S^{\rm MSTV} (\omega, \vec{q}, N=Z)
\simeq \frac{\Sigma ( m^2_\pi -\omega^2 +2\vec{q}^{\,2})}{f^2_\pi m^2_\pi}n\label{PiMSTV}\\
&\simeq\Pi_S^{\rm GSS} (\omega, \vec{q}, N=Z)\nonumber
\,,
\end{eqnarray}
which coincides with that used in \cite{Migdal:1990vm}, cf. also \cite{Jido2008},
and
\begin{eqnarray}
\delta \Pi_P^{\rm MSTV}\simeq -\gamma \vec{q}\vec{q}^{\,\prime}n\,.\label{deltaPMSTV}
 \end{eqnarray}
The value $\gamma$ can be constrained   from the analysis of the p-wave $\pi N$ scattering  amplitude and from the data on pionic atoms.  References \cite{Troitsky:1981,Migdal:1990vm,Voskresensky:1993ud}
used  $\delta \Pi_P^{\rm MSTV}=0$ fitting the parameters of the p-wave $\pi N$ interaction  from the analysis of the data on pion atoms.
If one takes $\gamma=-2c_3/f^2_\pi$,
then one gets $\delta \Pi_P^{\rm MSTV}=\delta \Pi_P^{\rm MG}=\delta \Pi_P^{\rm GSS}$.
The  condition (\ref{A2}) is satisfied for $\gamma =\Sigma/(f^2_\pi m^2_\pi)$.

 Constructing the
pion polarization operator in the gas approximation Refs. \cite{Delorme92,Ericson94,Kolomeitsev:2002pg}  conjectured to put in (\ref{DofDEE}) $q^2=q^{\prime\,2}=m^2_\pi$, exploiting that the amplitude of the $\pi N$ scattering in vacuum has physical sense  only for $q^2=q^{\prime\,2}=m^2_\pi$. In their approach  the amplitude (\ref{DofDEE})  satisfies  the condition that $a^+_{\pi N}\simeq 0$ and the Cheng-Dashen condition (\ref{ChengDashen}), whereas Weinberg and Adler conditions, as well as condition (\ref{A2}),  are not fulfilled. Then from (\ref{ChDMassSh}) one finds
\begin{eqnarray}
\widetilde{D}^+ (\omega^2,q^2=q^{\prime\,2}=m^2_\pi,\vec{q}\vec{q}^{\,\prime})\simeq  \frac{\Sigma (m^2_\pi -\omega^2)}{f^2_\pi m^2_\pi}+\gamma \vec{q}\vec{q}^{\,\prime}\,,\label{DofDEE2}
\end{eqnarray}
and
\begin{eqnarray}
&\Pi_S^{\rm KKW} (\omega, \vec{q}, N=Z)
\simeq \frac{\Sigma ( \omega^2 -m^2_\pi )}{f^2_\pi m^2_\pi}n\label{PiKKW}\\
&\simeq\Pi_S^{\rm MG} (\omega, \vec{q}, N=Z)\nonumber
\,,
\end{eqnarray}
$\delta \Pi_P$ remains the same as in Eq. (\ref{deltaPMSTV}).

As we have shown,  employment of the conjecture of \cite{Delorme92,Ericson94,Kolomeitsev:2002pg} resulting in expression (\ref{PiKKW})
does not work in the SM and GSS models but it works in the case of the MG model described by the reduced Lagrangian (\ref{MG}), where the off-mass-shell amplitude, as it follows from (\ref{MG}),  coincides with the on-mass-shell one. Thus, if one recovers the s-wave pion polarization operator employing   the scattering amplitude, rather than the Lagrangian, one cannot say   is it better to use  the  off-mass-shell amplitude or   the on-mass-shell one. Only after the Lagrangian of the model is  selected and approximation scheme  is chosen,   arbitrariness in the choice of the amplitude and the polarization operator disappears. They  follow directly from the model, although depending on the approximation employed for their calculation. In our examples above, we presented  $\Pi_S$ using the gas approximation.  In other cases one can use  perturbative expansions up to the given order \cite{IZ}, semiclassical series in the number of loops \cite{Kapusta}, expansion in number of vertices  in $\Phi$ derivable models \cite{IKV1}, expansion in classical field, as in Ginzburg-Landau model of phase transitions, etc. Confronting various physical effects  to the data one may then do a choice in favor of one model relatively others.

\section{S-wave pion condensation: to be or not to be?}\label{Tobe}
Using  Eq. (\ref{PiMSTV})  we may recover the corresponding effective pion Lagrangian density, written in the time-space representation.
Present it explicitly for the case of the spatially uniform charged pion field for simplicity:
\begin{eqnarray}
&L_{\rm ef}^{\rm MSTV}(\nabla\phi =0,N=Z)=|\dot{\phi}|^2 (1+{n}/{n_c})+\beta_1 |\ddot{\phi}|^2\nonumber\\
&-m^2_\pi |\phi|^2 (1+{n}/{n_c})-{\Lambda |\phi|^4}/{2}\,,\label{LMSTV}
\end{eqnarray}
where $n_c =f^2_\pi m^2_\pi/\Sigma$  and we recovered a small term $\propto \beta_1 >0$, the last term is responsible for the pion-pion effective interaction and  for simplicity we put  $\Lambda =const>0$. Actually, in the medium $\Lambda =\Lambda (\omega,\vec{q})$. As a typical value, we may take  $\Lambda\sim 1$, cf. \cite{Migdal:1978az}. Employing Eq. (\ref{PiKKW}) we get
\begin{eqnarray}
&L_{\rm ef}^{\rm KKW}(\nabla\phi =0,N=Z)=|\dot{\phi}|^2 (1-{n}/{n_c})+\beta_1 |\ddot{\phi}|^2\nonumber\\
&-m^2_\pi |\phi|^2 (1-{n}/{n_c})-{\Lambda |\phi|^4}/{2}\,.\label{KKW}
\end{eqnarray}

With $\phi =f e^{-i\omega t}$, where $f$ is the real constant, we recover
the energy densities
\begin{eqnarray}
&E_{\rm ef}^{\rm MSTV}(\nabla\phi =0,N=Z)=[\omega^2 (1+{n}/{n_c})+3\beta_1 \omega^4]\nonumber f^2 \\
&+m^2_\pi  (1+{n}/{n_c})f^2 +{\Lambda f^4}/{2}\,,
\end{eqnarray}
and
\begin{eqnarray}
&E_{\rm ef}^{\rm KKW}(\nabla\phi =0, N=Z)=[\omega^2 (1-{n}/{n_c})+3\beta_1 \omega^4]\nonumber f^2 \\
&+m^2_\pi  (1-{n}/{n_c})f^2 +{\Lambda f^4}/{2}\,.
\end{eqnarray}
As it is seen, $E_{\rm ef}^{\rm MSTV}(\nabla\phi =0)$ has minimum for $f=0$, i.e., s-wave pion condensation in isospin-symmetric matter does not occur, whereas $E_{\rm ef}^{\rm KKW}(\nabla\phi =0)<0$ for $n>n_c$. Thus in the latter model the s-wave pion condensation appears for $n>n_c$ (certainly, provided contributions beyong the gas approximation remain small up to $n\sim n_c$).

Further we focus on the KKW model, which allows for  the s-wave pion condensation for $n>n_c$. Minimization of the energy density in $\omega$ gives
\begin{eqnarray}\omega_m^2 =(n/n_c -1)/(6\beta_1)\,,
\end{eqnarray}
and
equation of motion,  $dL/d\phi =0$, yields
\begin{eqnarray}
&f^2 =[(\omega_m^2 +m^2_\pi)(n/n_c -1) -\beta_1 \omega_m^4)]\theta (n-n_c)/\Lambda
\nonumber\\
&\simeq m^2_\pi(n/n_c -1)\theta (n-n_c)/\Lambda+ O((n-n_c)^2)\,,\label{solution}
\end{eqnarray}
$\theta(x)$ is the step function.
Thus at least in the vicinity of the critical point the occurring pion field is quasistatic. Notice here that a  non-static complex field  carries  electric charge that modifies the initial $N/Z$ ratio. However for $n$ in the vicinity of the critical point the accumulated charge is only tiny and can be neglected. Setting solution (\ref{solution})  back to the energy density, we find
\begin{eqnarray}
&E_{\rm ef}^{\rm KKW}(N=Z)\simeq -\frac{m^4_\pi (n/n_c-1)^2}{2\Lambda}\theta (n-n_c)\\
&+O((n-n_c)^4)<0 \,.\nonumber
\end{eqnarray}
Thereby in the KKW model the s-wave pion condensation occurs in the isospin-symmetric nuclear matter by the second-order phase transition  at $n>n_c\simeq (2-2.5)n_0$ for $\Sigma\simeq 50-60$MeV. If one assumes $f_\pi (n)/f_\pi \simeq 1-0.1n/n_0$, one gets  $n_c \simeq (1.4- 1.7)n_0$. Note that employing the Gell-Mann-Oakes-Renner relation \cite{Gell-Mann68} $f^2_\pi (n)/f^2_\pi =1-\Sigma n/(m^2_\pi f^2_\pi)$ one would obtain a  still smaller value $f_\pi (n)/f_\pi\simeq 1-0.18n/n_0$ yielding a  smaller value of  $n_c$.

For $n/n_c \simeq 1.3$ we estimate the energy gain per particle due to s-wave pion condensation to be $-{\cal{E}}_{\rm ef}=-E_{\rm ef}/n\sim 0.1 m_\pi$.
In the non-linear Weinberg model \cite{Weinberg2009} for the case of the static field we estimate a stronger energy gain,
${\cal{E}}_{\rm ef}=-(n-n_c)\theta(n-n_c)m^2_\pi f^2_\pi/(n_c n)$ for $|\phi|^2= 2f^2_\pi$.
Estimated energy gains  could be sufficient for formation of  metastable (or may be even stable) s-wave pion condensate droplets already in heavy-ion collisions with energies $\lsim $ GeV$\cdot A$. In case of  the p-wave pion condensation such possibilities have been discussed in 1970s-1980s, cf. \cite{Migdal:1978az,Migdal:1990vm}.

Equation of motion for the time-dependent classical field $\phi$ renders
\begin{eqnarray}
(1- {n}/{n_c}) \ddot{\phi}-\beta_1 \ddddot{\phi}+ (1- {n}/{n_c})m^2_\pi \phi +\Lambda |\phi|^2\phi =0\,.\nonumber
\end{eqnarray}
For a slow field, dropping small term $\propto \beta_1$ we find a partial solution
\begin{eqnarray}
\phi (t)=e^{i\alpha}\,\theta (n-n_c)\sqrt{\frac{m^2 (n/n_c -1)}{\Lambda}}\mbox{th}\frac{m^2}{\sqrt{2}}t
\,,
\end{eqnarray}
where $\alpha$ is arbitrary constant. The states with different $\alpha$ are degenerate. In presence of  the  interaction term $\delta L=\epsilon_4 \phi\sqrt{\phi^*/\phi}\, +c.c.$, for a real value $\epsilon_4$, one would deal with the first order phase transition permitting  metastable and stable states.

Since  $n$ is an independent variable,
the quantities $\Pi_S (n)$, which we have estimated, may only slightly change with the temperature, $T$,  in a broad range of  the temperatures, cf. \cite{Voskresensky:1982vd,Migdal:1990vm}. Indeed, within the gas approximation  the $T$ dependence enters $\Pi_S$  via $f_\pi (T)$ and $\Sigma (T)$ and it becomes essential  only in the vicinity of the critical point of the deconfinement phase transition.
Thus, if the KKW model were valid, the s-wave pion condensation in the isospin-symmetric nuclear matter would be expected to occur already for $n>n_c\sim (1.4-2.5)n_0$ and one could expect to observe  some experimental consequences of the s-wave pion condensation in heavy-ion collisions in this case. Oppositely, with MSTV model for the s-wave pion-nucleon interaction the s-wave pion condensation in the isospin-symmetric nuclear matter does not occur at least up to very high densities.
\\

\section{Conclusion}\label{Conclusion}
We studied subtleties of the description of the s-wave pion-nucleon interaction in the isospin-symmetric nuclear matter. First, properties of the s-wave pion-nucleon interaction were studied on explicit examples of the linear sigma model, Lagrangian (\ref{linsigm1}), and   the Manohar-Georgi and  Gasser-Sainio-Svarc   models with finite number of terms in the  Lagrangians (reduced Lagrangians  (\ref{MG}) and (\ref{cutMG})  at $\alpha =\Sigma/(f^2_\pi m^2_\pi)$, respectively, provided terms labeled by dots in (\ref{MG}) are dropped).  On examples of the linear sigma and Gasser-Sainio-Svarc models we showed that the  knowledge of the  scattering amplitude, as a function of only $\nu$ variable,  for    $q^2=q^{\prime\,2}=m^2_\pi$, is  not sufficient to correctly describe the s-wave part of the pion polarization in matter even at low nucleon density. Even in the  lowest-order in $n$, the so called gas approximation, the s-wave pion polarization operator proves to be dependent on  the values of  $\nu$ and ${q}^{\,2}\neq m^2_\pi$  variables, as it straightly follows from the analysis of the Feynman diagrams (shown in Fig. \ref{PiNAmpl} in case of the sigma model)  and  the vertices of the Lagrangians. The key point here is that the gas approximation is applicable beyond the framework of the approximation of the scattering of free pions on static nucleon centers. Even if   complicated many-particle  processes  occur with only a  small probability (for low $n$), to calculate  the probability of such processes one requires the knowledge of the off-mass-shell pion-nucleon amplitude and the pion polarization operator for ${q}^{\,2}\neq m^2_\pi$.
Off-mass shell information is needed  for
the description of the zero-sound modes and the Landau damping in Fermi systems, even at low densities. Certainly, description of the phase transition phenomena at a higher density (clustering, Pomeranchuk instability, liquid-gas transition, etc, cf. \cite{Kolomeitsev2016}) also requires the knowledge of the in-medium scattering amplitudes and the dressed  Green functions.  More generally, particles are permanently produced and absorbed in the medium and do not exist in asymptotically free states, cf. \cite{IKV1,KV1996}. Thereby knowledge of the on-mass-shell amplitudes is not sufficient to calculate relevant physical quantities in all mentioned cases.

It does not contradict to the well known  equivalence theorem that  any  local change of variables in quantum field theories, which leaves the free field part of the Lagrangian unchanged, does not alter the $S$-matrix \cite{Chisholm1961}.  Coming back to  the problem considered in this paper, nucleons in matter undergo recoil effects (even, if  being small at low $n$) and do not fulfill the free Dirac equation and the pion field (especially the classical condensate field) does not obey the free Klein-Gordon equation. Thus,  even in the gas approximation (when effects nonlinear in $n$, although exist, are  small) the pion polarization operator is determined by the off-mass-shell pion-nucleon scattering amplitude rather than by the on-mass-shell one. Thereby, only selection  of the Lagrangian of the model allows one to determine unambiguously within the given  model and at the given approximation level  the physically important quantities. Only in the case of the Manohar-Georgi model  determined by the reduced Lagrangian (\ref{MG}) (when terms labeled by dots are dropped)    from those models we considered,  the s-wave pion polarization operator in the gas approximation proved to be independent on whether one uses off-shell or on-shell pion-nucleon scattering amplitudes.  In other cases for $q^2=q^{\prime\,2}=m^2_\pi$ (at arbitrary $\nu$), and for $q^2=q^{\prime\,2}\neq m^2_\pi$ even the pion spectra prove to be different in the non-linear density dependent terms.

We demonstrated that  the s-wave pion condensation in the isospin-symmetric matter hardly occurs within the linear sigma model and  the model described by the reduced Gasser-Sainio-Svarc Lagrangian, whereas it could occur in these models, if one artificially used the on-mass-shell amplitude to construct the pion polarization operator. However  within the model described by the reduced  Manohar-Georgi  Lagrangian (\ref{MG}), where the pion-nucleon amplitude (\ref{DMGappr}) does not depend explicitly  on values of $q^2$ and $q^{\prime\,2}$,   the s-wave pion condensation in the isospin-symmetric matter may appear already at $n>(2-2.5)n_0$ or even   for $n>1.4 n_0$, in the latter case provided the effective  pion decay constant  $f^*_\pi$ is decreased with increasing $n$.   Occurrence of the condensation   could result in appearance of  metastable (or may be even stable) condensate droplets already in heavy-ion collisions with energies $\lsim \, $GeV$\cdot A$. Further experimental check of presence or absence of these phenomena could help one to choose between employment of those phenomenological Lagrangians in the given problem. Important information on the pion polarization operator can be found from further  studies of deeply bound states in the single-pion and   double-pion atoms, cf.  \cite{Jido2021} and refs. therein.

In evaluation of the s-wave contribution to the pion polarization operator we disregarded correlation effects (we put $B=0$ in Eq. (\ref{Ps})) presenting only intuitive  arguments and mentioning rough estimates in favor of  their smallness.  However the quantitative consideration should be still done.

\begin{table}
\begin{tabular}{lllll}
\hline\hline
           & SM1      & SM2      & MG       & GSS\\
\hline
W (\ref{Weinberg})      & $+$      & $-$      & $-$      & $+$     \\
A(\ref{Adler})       & $+$      & $-$      & $-$      & $+$     \\
CD (\ref{ChengDashen})     & $+$      & $+$      & $+$      & $+$     \\
$\pi$, off & $-$ (\ref{PiSPof}) & $-$ (\ref{PSP2of}) & $+$ (\ref{PiMG}) & $-$ (\ref{PiGSS})\\
$\pi$, on  & $+$ (\ref{on}) & $+$ (\ref{Pon}) & $+$ (\ref{PiMG}) & $+$ (\ref{PiMG})\\
\hline\hline
\end{tabular}
\caption{Predictions on s-wave condensation in the isospin-symmetric matter in the SM1,2, reduced MG and reduced GSS models}
\label{tab:1}
\end{table}

In conclusion of this analysis, predictions on presence or absence of the s-wave condensation in the isospin-symmetric matter in the SM, reduced MG and reduced GSS models are illustrated in Table I, line $\pi$,off. The Weinberg, Adler and Cheng-Dashen conditions are fulfilled within SM1 and GSS models, whereas in SM2 and MG models only the Cheng-Dashen condition is fulfilled. In the full off-mass-shell treatment the s-wave pion condensation in isospin-symmetric matter may occur at $n\sim (1.4-2.5) n_0$ only in the reduced MG model. In the latter model the off-mass shell and on-mass shell treatments coincide. In the  artificial   models using the on-mass shell description, cf. line $\pi$,on, the s-wave pion condensation in isospin-symmetric matter may occur at $n\sim (1.4-2.5)n_0$ in all considered models.

Further, within a general phenomenological description not focusing on a specific model we constructed the fully off-mass-shell pion-nucleon scattering amplitude fitting parameters to satisfy the current algebra theorems and incorporating   smallness of the s-wave pion-nucleon  scattering length. The s-wave pion condensation in the isospin-symmetric matter hardly occurs in this model but it may occur already for $n>(1.4-2.5)n_0$,  provided one constructs the pion polarization operator in the gas approximation employing the on-mass-shell pion-nucleon  scattering amplitude and the fact of the smallness of the s-wave pion-nucleon  scattering length. As we have mentioned, the latter  procedure  works  in case  of the    model described by the  reduced Manohar-Georgi Lagrangian (\ref{MG}), but not  in cases of the linear sigma model and the GSS model described by the reduced Lagrangian $L_\alpha$ with $\alpha =\Sigma/(f^2_\pi m^2_\pi)$, cf.  (\ref{cutMG}).
Manifestation or non-manifestation  of  effects of the s-wave pion condensate  in   experimental investigations of  various nuclear systems (atomic nuclei, heavy-ion collisions, neutron stars) could  help to distinguish between different models.

Predictions on presence or absence of the s-wave condensation in the isospin-symmetric matter within the phenomenological  (Ph) expansion (\ref{DofDEE}) are illustrated in Table II, line ``Ph, off''. The Weinberg, Adler and Cheng-Dashen conditions are fulfilled and the s-wave pion condensation in isospin-symmetric matter does not occur at least up to a high density. Oppositely, in the on-mass shell treatment, line ``Ph, on'', only Cheng-Dashen condition is fulfilled and the s-wave pion condensation in isospin-symmetric matter may occur already  at $n\sim (1.4-2.5)n_0$.

\begin{table}[h]
\begin{tabular}{lllll}
\hline\hline
        & W   & A   & CD  & $\pi$-cond \\
\hline
Ph, off & $+$ & $+$ & $+$ & $-$ (\ref{PiMSTV}) \\
Ph, on  & $-$ & $-$ & $+$ & $+$ (\ref{PiKKW})\\
\hline\hline
\end{tabular}
\caption{Predictions on s-wave pion condensation in the isospin-symmetric matter in phenomenological description}
\label{tab:2}
\end{table}

We focused on the study of the isospin-symmetric matter, whereas consideration of the s-wave interaction in  asymmetric medium is   straightforward. For that, as a minimal step,  it is sufficient to incorporate the Weinberg-Tomozawa term.

In consideration of the kaon polarization in the matter  most of the authors focused on the phenomenon of the s-wave kaon condensation, whereas \cite{Kolomeitsev:1995xz,Kolomeitsev:1996bh,Kolomeitsev:2002pg} considered possibilities of both s- and p-wave condensations. Many works treated this problem  within the relativistic mean field models, cf. \cite{Glendenning2001}.
References \cite{Kolomeitsev:1995xz,Kolomeitsev:1996bh}
employed the low-energy theorems. Many other works, cf. \cite{AdamiBrown93,Kolomeitsev:2002pg},  used the on-mass-shell realization of the s-wave kaon-nucleon amplitude in matter putting $q^2=q^{\prime\,2}=m^2_K$.
All our caveats concerning subtleties of   the question about the s-wave pion polarization and condensation hold also for the case of the s-wave kaon polarization and condensation.

\acknowledgments I thank E.E. Kolomeitsev for numerous discussions and constructive critics. Discussions with Yu. B. Ivanov and
K. A. Maslov are also acknowledged.


\end{document}